\documentclass[longauth]{aa}
\usepackage{graphicx}
\usepackage{txfonts}
\usepackage{natbib}
\bibpunct{(}{)}{,}{a}{}{,}
\begin{document}
\newcommand{\doceCO}{\mbox{$^{12}$CO}}
\newcommand{\doce}{\mbox{$^{12}$CO}}
\newcommand{\trece}{\mbox{$^{13}$CO}}
\newcommand{\treceCO}{\mbox{$^{13}$CO}}
\newcommand{\dieciocho}{C\mbox{$^{18}$O}}
\newcommand{\jdn}{\mbox{$J$=10$-$9}}
\newcommand{\jdsq}{\mbox{$J$=16$-$15}}
\newcommand{\jsc}{\mbox{$J$=6$-$5}}
\newcommand{\jtd}{\mbox{$J$=3$-$2}}
\newcommand{\jcc}{\mbox{$J$=5$-$4}}
\newcommand{\jdu}{\mbox{$J$=2$-$1}}
\newcommand{\juc}{\mbox{$J$=1$-$0}}
\newcommand{\gsim}{\raisebox{-.4ex}{$\stackrel{>}{\scriptstyle \sim}$}}
\newcommand{\lsim}{\raisebox{-.4ex}{$\stackrel{<}{\scriptstyle \sim}$}}
\newcommand{\psim}{\raisebox{-.4ex}{$\stackrel{\propto}{\scriptstyle \sim}$}}
\newcommand{\kms}{\mbox{km~s$^{-1}$}}
\newcommand{\s}{\mbox{$''$}}
\newcommand{\mloss}{\mbox{$\dot{M}$}}
\newcommand{\my}{\mbox{$M_{\odot}$~yr$^{-1}$}}
\newcommand{\ls}{\mbox{$L_{\odot}$}}
\newcommand{\ms}{\mbox{$M_{\odot}$}}
\newcommand{\mm}{\mbox{$\mu$m}}
\def\arcdeg{\hbox{$^\circ$}}
\newcommand{\seca}{\mbox{\rlap{.}$''$}}
\newcommand{\dega}{\mbox{\rlap{.}$^\circ$}}
\newcommand{\aprop}{\raisebox{-.4ex}{$\stackrel{\propto}{\scriptstyle\sf \sim}$}}
\newcommand{\apropg}{\raisebox{-.4ex}{$\stackrel{\Large \propto}{\sim}$}}
\title{Herschel/HIFI observations of molecular emission in 
  protoplanetary nebulae and young planetary nebulae\thanks{Herschel is an ESA space
  observatory with science instruments provided by European-led
  Principal Investigator consortia and with important participation
  from NASA.}}

   \subtitle{}

   \author{V.~Bujarrabal
          \inst{1}
          \and
J.~Alcolea
          \inst{2}
\and
R.~Soria-Ruiz
\inst{2}
\and
P.~Planesas
\inst{2,12}
\and
D.~Teyssier
          \inst{3}
\and
J.~Cernicharo
\inst{4}
\and
L.~Decin
\inst{5,6}
\and
C.~Dominik
\inst{6,13}
\and
K.~Justtanont
\inst{7}
\and
A.~de~Koter
\inst{6,14}
\and
A.P.~Marston
\inst{3}
\and
G.~Melnick
\inst{8}
\and
K.M.~Menten
\inst{9}
\and
D.A.~Neufeld
\inst{10}
\and
H.~Olofsson
\inst{7,15}
     \and 
M.~Schmidt
\inst{11}
\and
F.L.~Sch\"oier
\inst{7}
     \and 
R.~Szczerba
\inst{11}
\and
L.B.F.M.~Waters
\inst{6,5}
          }


   \institute{   
Observatorio Astron\'omico Nacional (IGN), Ap 112, E--28803 
Alcal\'a de Henares, Spain\\
              \email{v.bujarrabal@oan.es}
	      \and
Observatorio Astron\'omico Nacional (IGN), Alfonso XII N$^{\circ}$3,
              E--28014 Madrid, Spain  
\and
European Space Astronomy Centre, ESA, P.O. Box 78, E--28691
Villanueva de la Ca\~nada, Madrid, Spain
\and
CAB, INTA-CSIC, 
Ctra de Torrej\'on a Ajalvir, km 4,
E--28850 Torrej\'on de Ardoz, Madrid, Spain
\and
Instituut voor Sterrenkunde,
             Katholieke Universiteit Leuven, Celestijnenlaan 200D, 3001
Leuven, Belgium
\and
    Sterrenkundig Instituut Anton Pannekoek, University of Amsterdam,
Science Park 904, NL-1098 Amsterdam, The Netherlands
\and
Onsala Space Observatory,  Dept. of Radio and Space Science, Chalmers  
University of Technology, SE--43992 Onsala, Sweden
\and
Harvard-Smithsonian Center for Astrophysics, Cambridge, MA 02138, USA
\and
Max-Planck-Institut f{\"u}r Radioastronomie, Auf dem H{\"u}gel 69,
D-53121 Bonn, Germany 
\and
The Johns Hopkins University, 3400 North Charles St, Baltimore, MD  
21218, USA
\and
N. Copernicus Astronomical Center, Rabia{\'n}ska 8, 87-100 Toru{\'n}, Poland
\and
Joint ALMA Observatory, El Golf 40, Las Condes, Santiago, Chile
\and 
Department of Astrophysics/IMAPP, Radboud University Nijmegen,   
Nijmegen, The Netherlands
\and
Astronomical Institute, Utrecht University,
Princetonplein 5, 3584 CC Utrecht, The Netherlands 
\and
Department of Astronomy, AlbaNova University Center, Stockholm  
University, SE--10691 Stockholm, Sweden
}

   \date{accepted}

 
  \abstract
   {}
   {We aim to study the physical conditions, particularly the
  excitation state, of the intermediate-temperature gas in 
 protoplanetary nebulae and young planetary nebulae (PPNe, PNe). 
The information that the observations of the different components deliver
  is of particular importance for understanding the evolution of these 
  objects.}
  {We performed Herschel/HIFI observations of intermediate-excitation 
  molecular lines in the far-infrared/submillimeter range in a sample
  of ten nebulae. The high spectral resolution provided by HIFI
  allows the accurate measurement of the line profiles. The
  dynamics and evolution of these nebulae are known to result from the
  presence of  
  several gas components, notably fast bipolar outflows and slow
  shells (that often are the fossil AGB shells), and the interaction
  between them. Because of the diverse
  kinematic properties of the different components, their emissions can
  be identified in the line profiles. The observation of
  these high-energy transitions allows an accurate study of the
  excitation conditions, particularly in the warm gas, 
which cannot be properly studied from the low-energy lines. 
  }
   {We have detected FIR/sub-mm lines of several molecules, in
  particular of \doce, \trece, and H$_2$O. Emission from other species,
  like NH$_3$, OH, H$_2$$^{18}$O, HCN, SiO, etc, has been also detected. 
  Wide profiles showing sometimes spectacular line wings have been
  found. We have mainly studied the excitation properties of the
  high-velocity emission, which is known to come from fast bipolar
  outflows. From comparison with general theoretical predictions, we
  find that CRL\,618 shows a particularly warm fast wind, with 
  characteristic kinetic temperature $T_{\rm k}$ \gsim\ 200 K. In
  contrast,  
  the fast winds in OH\,231.8+4.2 and NGC\,6302 are cold, $T_{\rm k}$
  $\sim$ 30 K. Other nebulae, like CRL\,2688, show intermediate
  temperatures, with characteristic values around 100 K.
  We also discuss how the complex structure of the nebulae can affect
  our estimates, considering two-component models. We argue
  that the differences in temperature in the different nebulae can be
  due to cooling after the gas acceleration (that is probably due to
  shocks); for instance, CRL\,618 is a case of very recent acceleration,
  less than $\sim$ 100 yr ago, while the fast gas in OH\,231.8+4.2 was
  accelerated $\sim$ 1000 yr ago. We also find indications that the
  densest gas tends to be cooler, which may be explained by the
  expected increase of the radiative
  cooling efficiency with the density. 
  }
{}

   \keywords{stars: AGB and post-AGB -- stars: circumstellar matter,
               mass-loss -- planetary nebulae}

   \maketitle
%

\section{Introduction}

Asymptotic giant branch (AGB) stars, together with the thick
circumstellar envelopes (CSEs) around them, evolve to form planetary
nebulae (PNe), through the phase of protoplanetary nebulae (PPNe). AGB
stars are mass-losing red giants surrounded by spherical, slowly
expanding CSEs, while PNe, surrounding blue or white dwarfs, show in
most cases strong axial symmetry or ring-like geometries and very fast
bipolar jets. The transition time between these very different stages
is extremely short, only about 1000 yr, during which the intermediate
PPNe develop a wide variety of structures.

PPNe and young PNe already present fast bipolar outflows, along with
slower components with velocities similar to those of AGB
envelopes. These bipolar flows typically reach velocities of 50--100
\kms, and affect a sizable fraction of the nebular mass, $\sim$
0.1--0.3 \ms\ \citep{bujetal01}.
Such remarkable dynamics is thought to be the result of the interaction
between the AGB and post-AGB winds: axial, very fast jets ejected in
the very first post-AGB phases would collide with the denser material
ejected during the AGB phase \citep[e.g.][]{balickf02}. The presently
observed bipolar outflows would then correspond to a part of the
relatively dense shells ejected during the last AGB phase, mostly their
polar regions, accelerated by the shocks that propagate
during the PPN phase. Many aspects of this process are, however, not
yet well understood.

The massive bipolar outflows in PPNe, as well as the unaltered remnants
of the AGB shells, usually show strong emission in molecular lines
\citep[e.g.][]{bujetal01}, which is the best tool to study these
nebulae. Several PPNe have been extensively observed in mm-wave lines,
including interferometric maps with resolutions $\sim$ 1$''$. Thanks to
those observations, the structure, dynamics, and physical conditions in
a number of PPNe are quite well known.  The profiles of the
molecular lines have been found to be composite, showing in most sources
a central core plus wide line wings. Maps indicate that the line core
basically comes from the remnant of the slow AGB wind, which is found
to be roughly
spherical or confined to equatorial regions. On the other hand, the
line wings come from the fast bipolar outflows, which are 
more or less elongated in the direction of the nebular axis and 
show in most sources a strong velocity
gradient, the velocity increasing proportionally to the distance to the
star. See examples in \cite{sanchezc04,fong06,alco07,ccarrizo10}, and
further references given below.  

However, observations of the low-$J$ transitions are not very useful
for studying the warm gas components, in particular those heated by the
shocks. The often observed \jdu\ and \juc\ transitions of CO only
require temperatures $T_{\rm k}$ $\sim$ 15 K to be excited. Indeed
their maximum emissivity occurs for excitation temperatures of 10--20
K, and the line intensities and line intensity ratios depend only
slightly on the excitation state in relatively warm gas. Of course,
observations in the visible or near infrared ranges tend to select hot
regions, with typical temperatures over 1000 K. The proper study of
warm gas, 100 K \lsim\ $T_{\rm k}$ \lsim\ 1000 K, therefore requires
observations at intermediate wavelengths, in the far infrared (FIR) and
sub-mm ranges.

Because of the crucial role of shocks in the formation of PNe, these
warm regions are particularly important for understanding the nebular
structure and evolution. In some well-studied cases, e.g.\ M\,1--92,
OH\,231.8+4.2, and M\,2--56 \citep{bujetal98,alco01,ccarrizo02}, the
high-velocity, massive outflows seem to be very cold, with temperatures
\lsim\ 20--30 K, which implies very fast cooling in the
shock-accelerated gas. No warm component, presumably heated by recent
shocks, has been identified in these sources until now.

The only PPN in which detailed studies of warm accelerated gas have
been peformed is CRL\,618. Interferometric imaging of the \doce\ \jdu\
line shows that dense gas in axial structures presents higher
temperatures, typically $\sim$ 50--100 K \citep[][]{sanchezc04}.  But,
precisely because of their relatively high excitation, the temperature
estimate in these components from CO \jdu\ is uncertain. Maps of
\doce\ \jsc\ were also obtained by Nakashima et al.\ (2007), but with
much less information on the nebular structure. A first attempt to
study such warm regions in the FIR was performed by
\cite{justtanont00}, from ISO measurements of CO lines as high as
$J$=37--36 in CRL\,618, CRL\,2688, and NGC\,7027. These authors indeed
found temperatures of several hundred degrees in these sources. The
lack of spectral resolution in the ISO data prevented any discussion on
the components responsible for the line emission, which made difficult
the interpretation of these results in nebulae showing very complex
structure and dynamics. (This work and other previous ones are
discussed in more detail in Sect.\ 3.2.) Recent Herschel/HIFI
observations in the FIR/sub-mm of CRL\,618 \citep[][]{bujetal10}, with
sufficiently high spectral resolution, have confirmed the presence of
high temperature in various kinematic components. The temperature
derived by \cite{sanchezc04} was found to be too low, the fast gas
showing temperatures typically $\sim$ 200 K in the bipolar outflow, and
significantly higher in its base. Similar high temperatures were found
in the innermost regions of the slow component.

The Herschel Space Telescope is well-suited for studying warm gas around
evolved stars in the FIR and sub-mm. The high spectral resolution that
can be achieved with the heterodyne instrument HIFI (better than
1\,\kms) is particularly important in our case: it is useful to identify
the different nebular components in the profiles and allows the analysis
of the kinematics, a fundamental parameter in the study of this warm,
shocked gas. Here we present Herschel/HIFI observations in a sample of
ten PPNe and young PNe of several intermediate-excitation lines of
abundant molecules: \doce\ and \trece, H$_2$O, OH, NH$_3$, HCN,
SiO, etc. These observations have been obtained as part of the Herschel
guaranteed-time key program HIFISTARS, which is devoted to the study of
intermediate-excitation molecular lines in nebulae around evolved
stars.

\begin{figure}
\vspace{0cm}
{\hspace{-0.1cm}\vspace{0cm}\resizebox{9cm}{!}{ 
\includegraphics{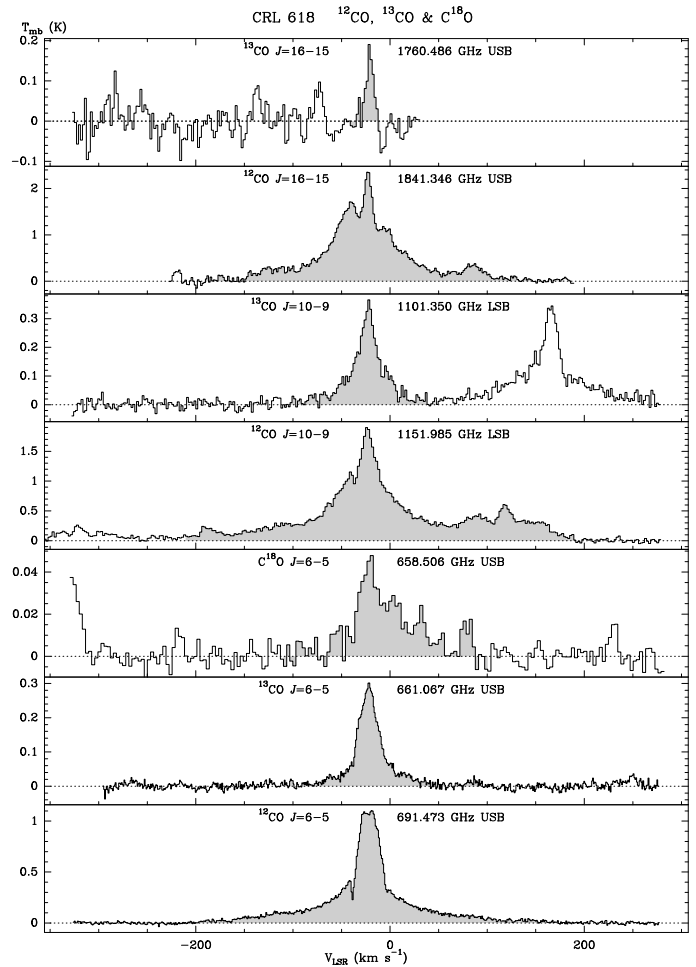}
}}
\caption{HIFI observations including detected \doce, \trece, and
  C$^{18}$O lines
  in CRL\,618 ($T_{\rm mb}$ vs.\ {\em LSR} velocities). The shadowed
  regions indicate the velocity ranges in which we can expect emission
  from the considered line, often mixed with emission from other
  lines. Note for instance the emission of HCN superposed to \doce\
  10--9 and the overlapping lines \doce\ 16--15 and OH 3/2--1/2 (for
  more details, see the observed full-band spectra in the Appendix).  }
\end{figure}

\begin{figure}
\vspace{-0cm}
{\hspace{-0.1cm}\vspace{-0cm}\resizebox{9cm}{!}{ 
\includegraphics{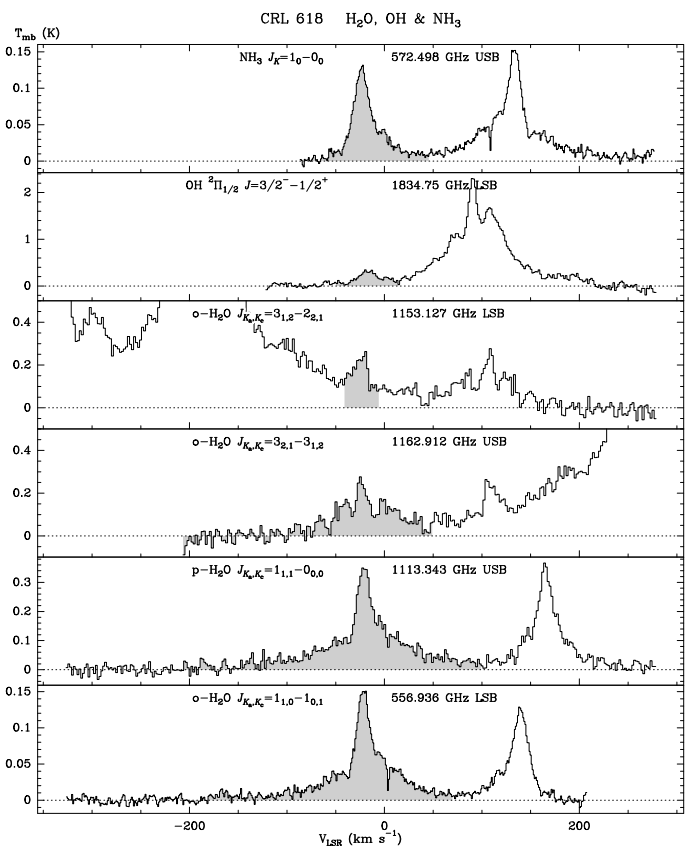}
}}
\caption{HIFI observations of CRL\,618 including detected lines of
  H$_2$O, OH, and NH$_3$ ($T_{\rm mb}$ vs.\ {\em LSR} velocities). The
  shadowed regions indicate the velocity ranges in which we can expect
  emission from the considered line, often mixed with emission from
  other lines (see identifications in Fig.\ A.1). }
\end{figure}

\begin{figure}
\vspace{0cm}
{\hspace{-0.1cm}\vspace{-0cm}\resizebox{9cm}{!}{ 
\includegraphics{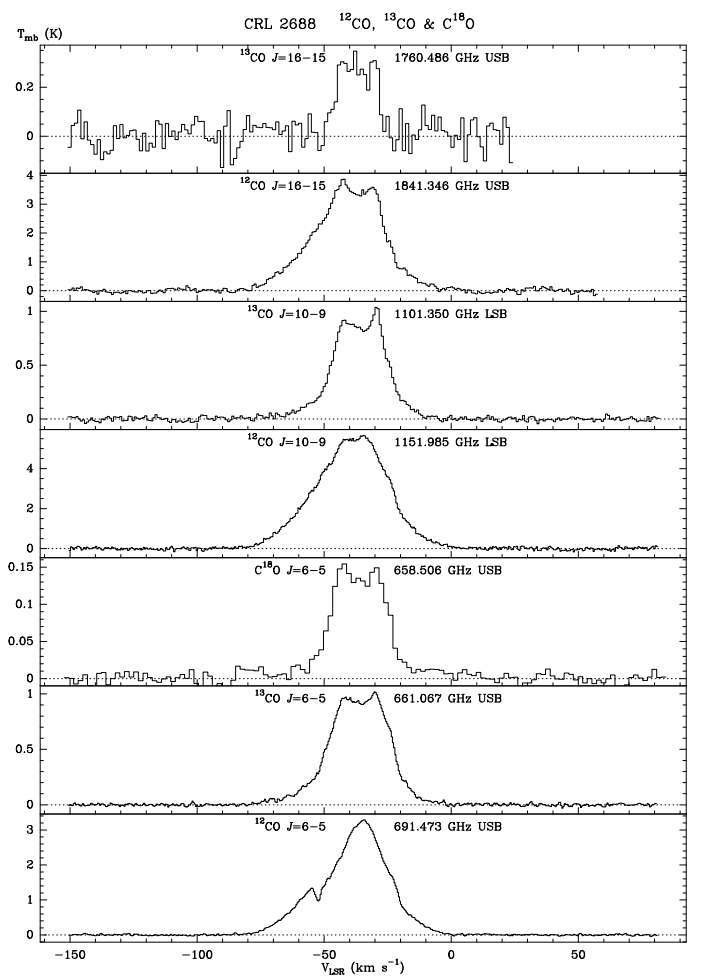}
}}
\caption{HIFI observations including detected \doce, \trece, and
  C$^{18}$O lines
  in CRL\,2688 ($T_{\rm mb}$ vs.\ {\em LSR} velocities).
  }
\end{figure}

\begin{figure}
\vspace{-.0cm}
{\hspace{-0.1cm}\vspace{-0cm}\resizebox{9cm}{!}{ 
\includegraphics{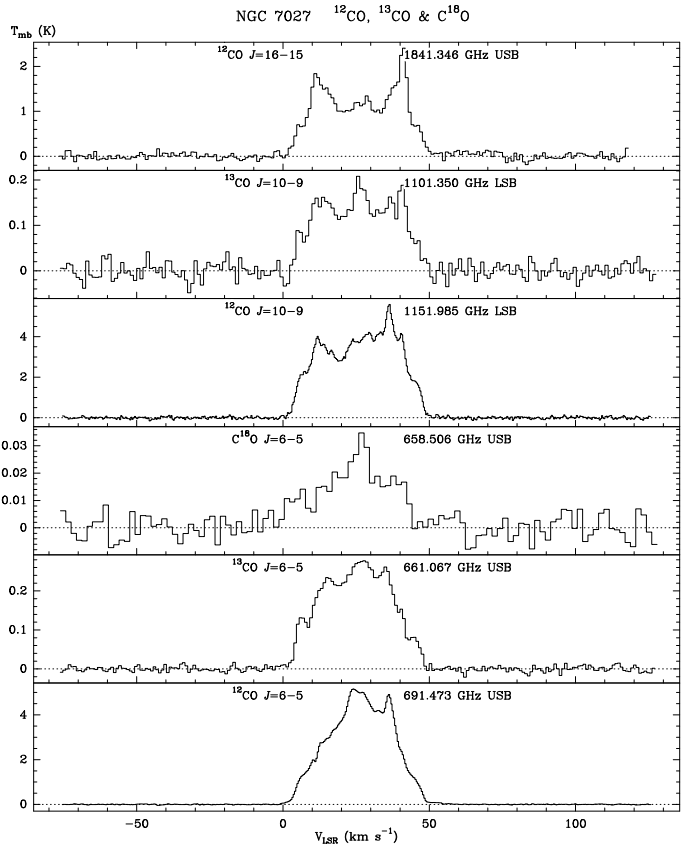}
}}
\caption{HIFI observations including detected \doce, \trece, and
  C$^{18}$O lines
  in NGC\,7027 ($T_{\rm mb}$ vs.\ {\em LSR} velocities).
} 
\end{figure}

\begin{figure}
\vspace{-0cm}
{\hspace{-.1cm}\vspace{-0cm}\resizebox{9cm}{!}{ 
\includegraphics{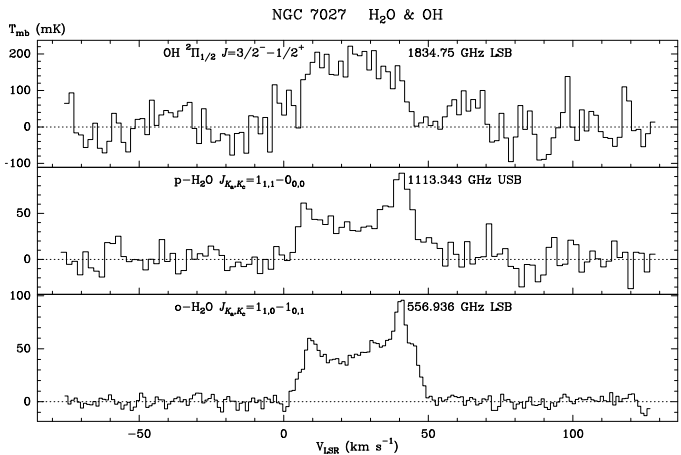}
}}
\caption{HIFI observations of NGC\,7027 including detected lines of
  H$_2$O and OH ($T_{\rm mb}$ vs.\ {\em LSR} velocities).
} 
\end{figure}

\begin{figure}
\vspace{0cm}
{\hspace{-0.1cm}\vspace{-0cm}\resizebox{9cm}{!}{ 
\includegraphics{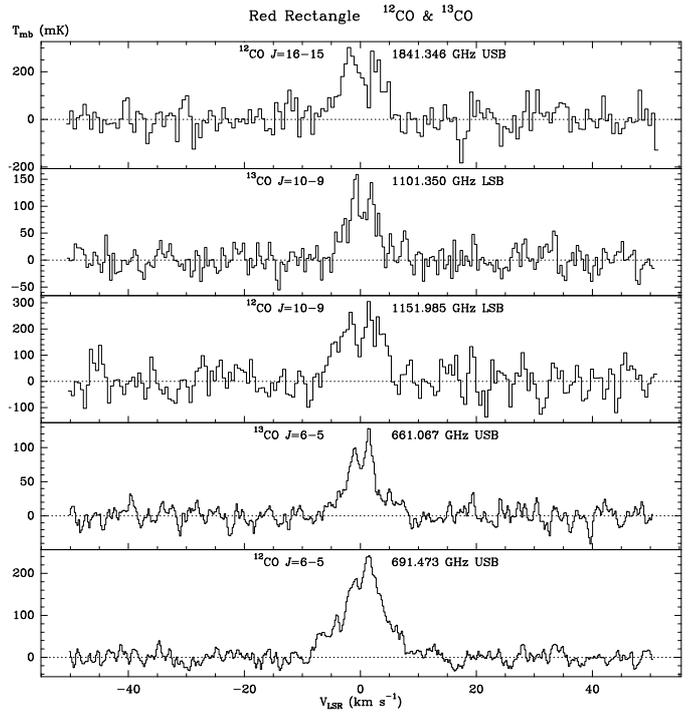}
}}
\caption{HIFI observations including detected \doce\ and \trece\ lines
  in the Red Rectangle ($T_{\rm mb}$ vs.\ {\em LSR} velocities).
  }
\end{figure}

\begin{figure}
\vspace{0cm}
{\hspace{-0.1cm}\vspace{-0cm}\resizebox{9cm}{!}{ 
\includegraphics{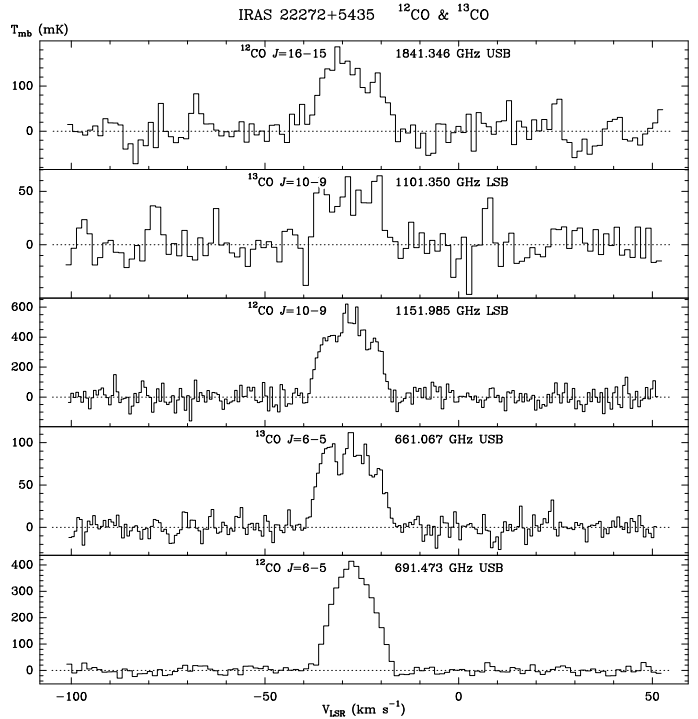}
}}
\caption{HIFI observations including detected \doce\ and \trece\ lines
  in IRAS\,22272+5435 ($T_{\rm mb}$ vs.\ {\em LSR} velocities).
  }
\end{figure}

\begin{figure}
\vspace{0.cm}
{\hspace{-0.1cm}\vspace{0cm}\resizebox{9cm}{!}{ 
\includegraphics{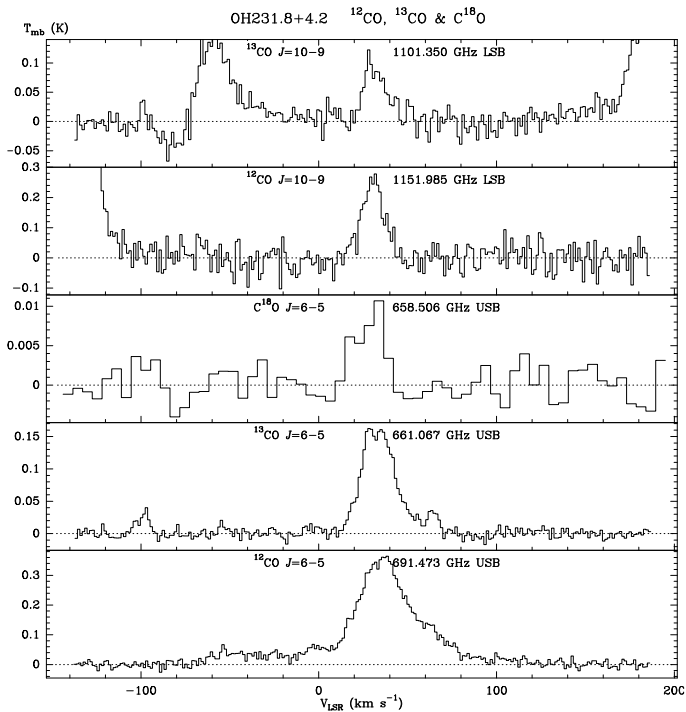}
}}
\caption{HIFI observations including detected \doce, \trece, and
  C$^{18}$O lines in OH\,231.8+4.2 ($T_{\rm mb}$ vs.\ {\em LSR}
  velocities). Note that other lines appear in some spectra, see Fig.\ A.6.}
\end{figure}

\begin{figure}
\vspace{-.0cm}
{\hspace{-0.1cm}\vspace{0cm}\resizebox{9cm}{!}{ 
\includegraphics{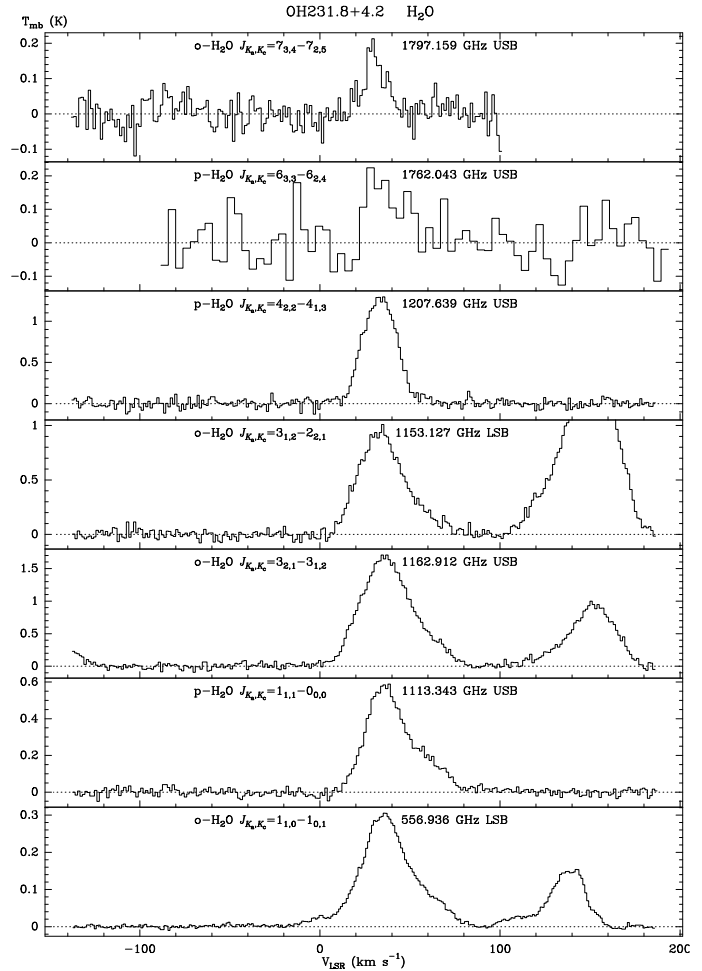}
}}
\caption{HIFI observations of OH\,231.8+4.2 including detected lines of
  H$_2$O ($T_{\rm mb}$ vs.\ {\em LSR} velocities). Note that other
  lines appear in some spectra, see Fig.\ A.6.  }
\end{figure}

\begin{figure}
\vspace{0cm}
{\hspace{-0.1cm}\vspace{-0cm}\resizebox{9cm}{!}{ 
\includegraphics{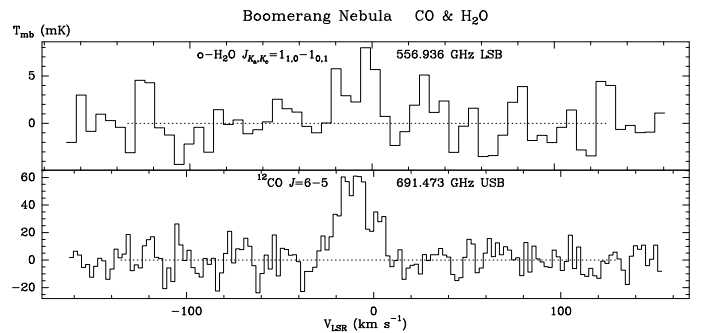}
}}
\caption{HIFI observations including detected \doce\ and H$_2$O lines
  in the Boomerang Nebula ($T_{\rm mb}$ vs.\ {\em LSR} velocities).
  }
\end{figure}

\begin{figure}
\vspace{-0cm}
{\hspace{-.1cm}\vspace{-0cm}\resizebox{9cm}{!}{ 
\includegraphics{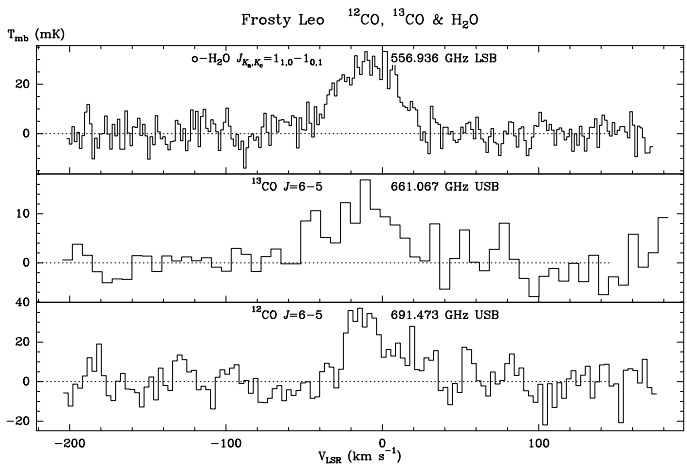}
}}
\caption{HIFI observations including detected \doce, \trece, and
  H$_2$O lines in Frosty Leo ($T_{\rm mb}$ vs.\ {\em LSR} velocities).
  }
\end{figure}

\begin{figure}
\vspace{-.0cm}
{\hspace{-0.1cm}\vspace{-0cm}\resizebox{9cm}{!}{ 
\includegraphics{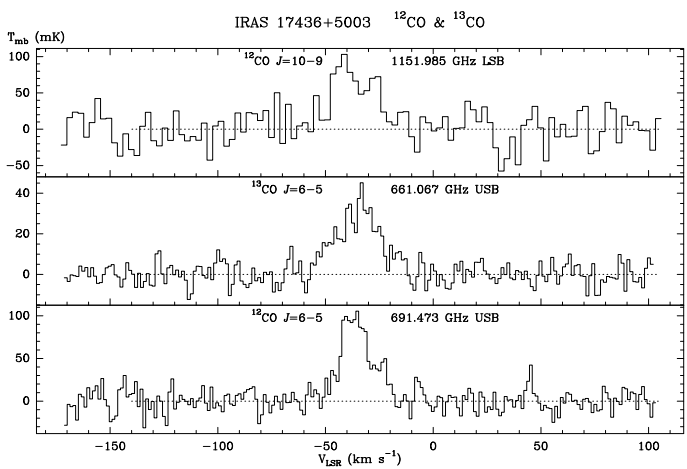}
}}
\caption{HIFI observations including detected \doce\ and \trece\ lines,
$T_{\rm mb}$ vs.\ {\em LSR} velocities, 
in IRAS\,17436+5003 (HD\,161796).
    }
\end{figure}

\begin{figure}
\vspace{-.0cm}
{\hspace{-0.1cm}\vspace{-0cm}\resizebox{9cm}{!}{ 
\includegraphics{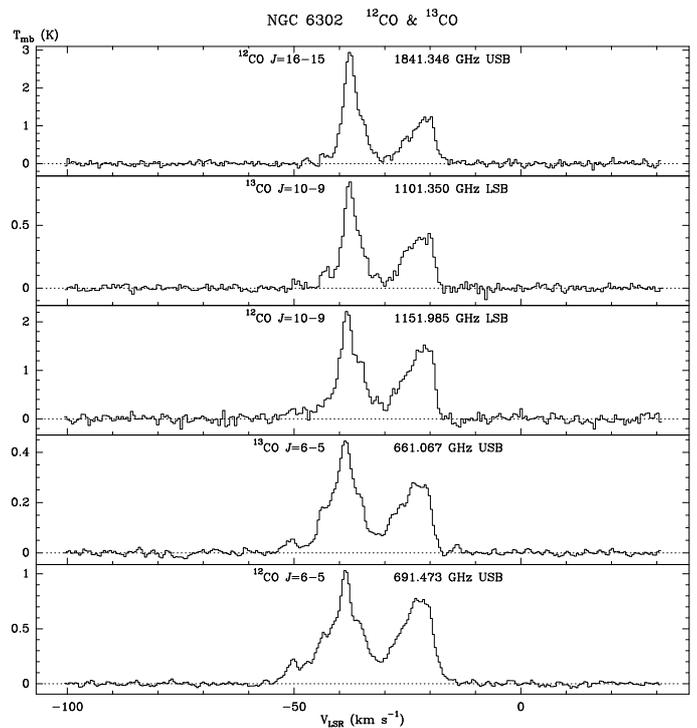}
}}
\caption{HIFI observations including detected \doce\ and \trece\ lines
  in NGC\,6302 ($T_{\rm mb}$ vs.\ {\em LSR} velocities).
    }
\end{figure}

\begin{figure}
\vspace{-.0cm}
{\hspace{-0.1cm}\vspace{-0cm}\resizebox{9cm}{!}{ 
\includegraphics{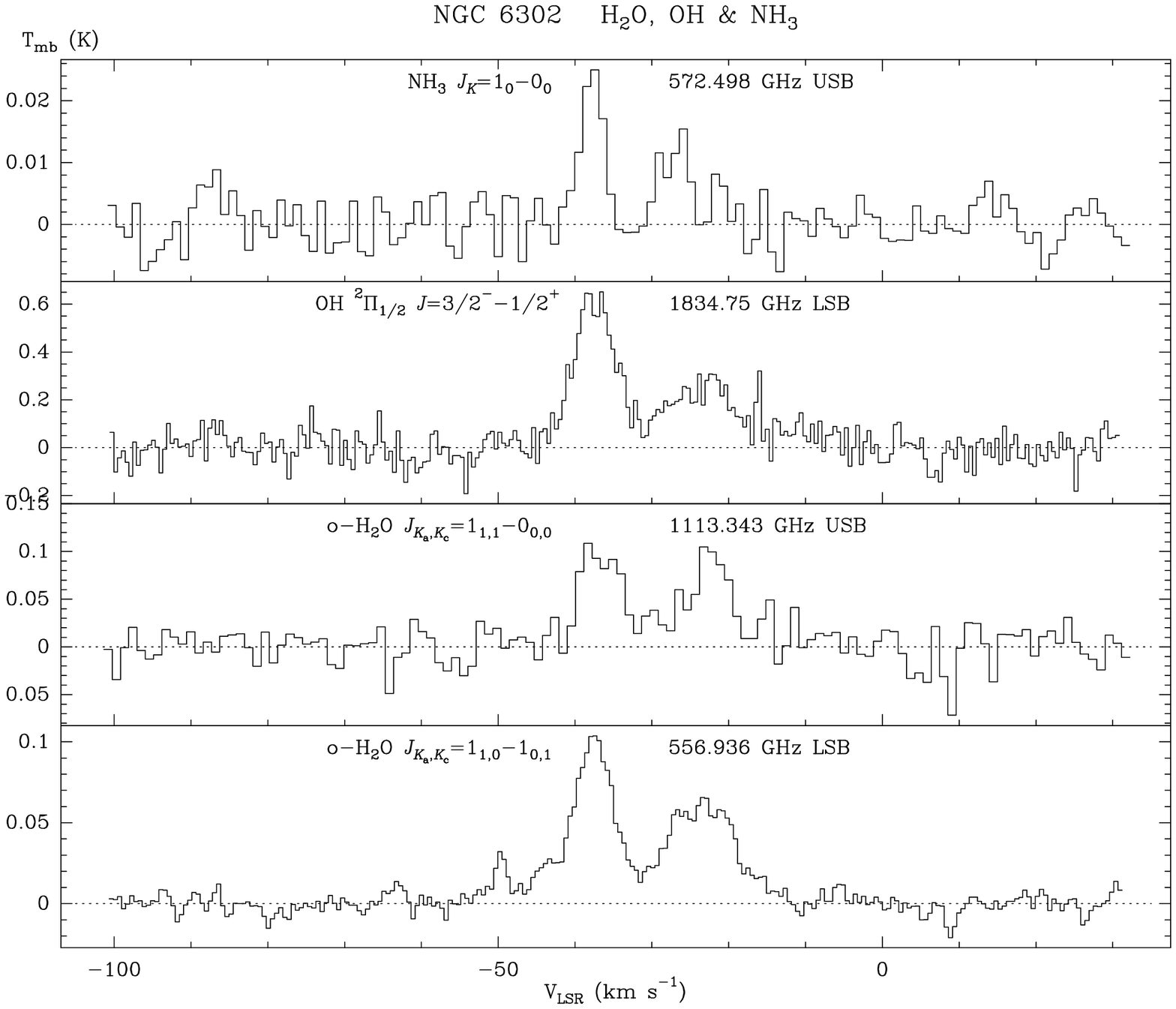}
}}
\caption{HIFI observations including detected lines of
  H$_2$O, OH, and NH$_3$ in NGC\,6302 ($T_{\rm mb}$ vs.\ {\em LSR}
  velocities). 
  }
\end{figure}

\section{Observations}

We used the HIFI heterodyne spectrometer, on board the Herschel Space
Telescope \citep{pilbratt10, degraauw10}, to observe a number of
molecular lines in a total of ten protoplanetary nebulae (PPNe). The
observed objects are given in Table 1, where we quote the observed
coordinates and additional properties of the sources.

\begin{table*}
\caption{Observed sources.}  
\begin{center}
\begin{tabular}{|l|r|r|c|l|l|}
\hline
source & $\alpha$ (J2000) & $\delta$ (J2000) & central star &
   previous CO data & comments; distance  \\
\hline
\multicolumn{6}{|l|}{\bf C-rich nebulae} \\
\hline
CRL\,618    &  04 42 53.7 & +36 06 53.2 & B0 &  2--1 and
1--0 maps$^1$, high-$J$ data$^2$ &   \\
CRL\,2688 & 21 02 18.7 & +36 41 37.8  & F5Iae & 2--1 maps$^3$ &  
 \\
NGC\,7027  &  21 07 01.6 & +42 14 10.2 & O & 2--1 and
1--0 maps$^{4}$ & high-excitation nebula \\
Red Rectangle  & 06 19 58.2 & $-$10 38 14.7 & A1 & 2--1 and
1--0 maps$^{5}$ & equatorial disk in rotation \\
IRAS\,22272+5435 & 22 29 10.4 & +54 51 06.3  & G5Ia & 
1--0 maps$^{6}$, 2--1 and 1--0 spectra$^{7}$ & isotrpic shell, no
bipolar flow \\
\hline
\hline
\multicolumn{6}{|l|}{\bf O-rich nebulae} \\
\hline
OH\,231.8+4.2 & 07 42 16.8 & $-$14 42 52.1 & M9III + AV & 2--1 and
1--0 maps$^8$, high-$J$ data$^9$ & Calabash Nebula  \\
Boomerang Nebula & 12 44 45.4 & $-$54 31 11.4 & G0III &  2--1 and
1--0 spectra$^{10}$ &  \\
Frosty Leo   &  9 39 54.0 & +11 58 54.0 & K7III & 2--1 and
1--0 maps$^{11}$  &  \\
IRAS\,17436+5003 & 17 44 55.4 & +50 02 39.5 & F2-5Ib & 2--1 and
1--0 spectra$^7$ & HD\,161796  \\
NGC\,6302 & 17 13 44.2 & $-$37 06 15.9 & O  & 2--1
maps$^{12}$ & high-excitation nebula \\
\hline
\end{tabular}
\end{center}
{\vspace{-.2cm}\bf References:} $^1$:
  \cite{sanchezc04}; $^2$: \cite{bujetal10,naka07}; $^3$: \cite{cox00};
  $^4$:  \cite{naka10}; $^5$: \cite{bujetal05}; $^6$: \cite{fong06};
  $^7$: \cite{bujetal01}; $^8$: \cite{alco01}; $^9$: \cite{alco11};
  $^{10}$: \cite{bujb91,sahai97}; $^{11}$: \cite{ccarrizo05}; 
$^{12}$: \cite{trung08}.
\end{table*}

Herschel/HIFI is an instrument very well suited to observe molecular
lines in the FIR and sub-mm domains, providing a very high spectral
resolution that allows the line profiles to be resolved.  The data were
taken using the two orthogonal HIFI receivers, H and V, available at
each band, which were systematically averaged after checking that there
was no significant difference in the lines detected in both
receivers. The spectrometer works in double side-band (DSB) mode, which
effectively doubles the instantaneous IF coverage. Care was taken when
choosing the local oscillator frequency, maximizing the number of
interesting lines that could be observed simultaneously.

A total of nine frequency settings were used (not all of them in all
sources), between 557 and 1843 GHz. The settings were chosen to observe
the \doce\ and \trece\ \jsc, \jdn, and \jdsq\ lines and the H$_2$O
1$_{1,0}$--1$_{0,1}$, 1$_{1,1}$--0$_{0,0}$, 3$_{2,1}$--3$_{1,2}$,
3$_{1,2}$--2$_{2,1}$, 4$_{2,2}$--4$_{1,3}$, and 7$_{3,4}$--7$_{2,5}$
lines. Due to the wide simultaneous frequency bands observable with
HIFI and the double side band system used, each frequency setting often 
include other lines. Interesting molecular transitions also
placed within the observed bands are the NH$_3$ $J,K$=1,0--0,0 and OH
$^2\Pi_{1/2}$~3/2--1/2 rotational lines, as well as lines of C$^{18}$O,
H$_2$$^{18}$O, vibrationally excited H$_2$O, SiO, HCN, CN, etc.

The observed full-band spectra are reproduced in the Appendix. In those
figures we indicate the upper and lower side band (USB and LSB)
frequency scales and the frequencies of interesting lines placed within
the observed bands. Continuum has been subtracted in all the spectra
presented in this paper.

The observations were obtained in the dual-beam-switching (DBS)
mode. In this mode, the HIFI internal steering mirror chops between the
source position and an emission-free position 3$'$ away. The telescope
then alternatively locates the source in either of the chopped beams,
providing a double-difference calibration scheme, which allows a more
efficient cancellation of the residual standing waves in the spectra. 
This procedure works very well except
for the highest frequencies (HIFI band 7, above 1700 GHz), where
noticeable ripples were found in some spectra, specially in the
V-receiver. Other spurious effects in the baselines were found in a few
cases, particularly around 1200 GHz due to very different behaviors of
the sub-band baselines, which were sometimes difficult to remove.

The HIFI data shown here were taken using the Wide-Band Spectrometer
(WBS), an acousto-optical spectrometer that provides simultaneous
coverage of the full instantaneous IF band in the two available
orthogonal receivers, with a spectral resolution of about 1.1 MHz. In
some cases, the spectra have been smoothed to improve the S/N ratio.

We processed the data using the standard HIFI pipeline in the HIPE
software, with a modified version of the level 2 algorithm that yields
unaveraged spectra with all spectrometer sub-bands stitched together.
Later on, the spectra were exported to CLASS using the hiClass tool
within HIPE, for further inspection, ``flagging'' data with outstanding
ripple residuals, final averaging and calibration, and baseline
removal. Even if spurious effects in the baselines were in a few cases
impossible to remove, we have only considered baselines of moderate
degree, in general just of degree 1, at most of degree 3.

The data were originally calibrated in antenna temperature units and
later converted into main-beam temperatures ($T_ {\rm mb}$). 
A summary of the telescope characteristics and observational
uncertainties, including the spatial resolution and the
conversion factors to main-beam units, can be found in the Herschel
Observers' Manual and the HIFI Instrument and Calibration Web Page (see
http://herschel.esac.esa.int).  We recall that the FWHM
beam sizes range between about 33 and 12 arcsec (inversely
proportional to the observed frequency) and $T_{\rm mb}$/$T_{\rm A}$
ratios range between 1.3 and 1.5.

In Figs.\ 1 to 14, we show zoom-ups of selected lines detected in our
sources, in {\em LSR} velocity and $T_{\rm mb}$ units. In CRL\,618,
some of these lines were already presented and discussed by
\cite{bujetal10}; here we have reanalyzed all the data and included new
observations.
In Figs.\ 1 and 2, corresponding to CRL\,618, we have indicated in grey
the velocity range expected for each line profile, because of the very
complex spectra and wide profiles characteristic of this nebula. Data
of the most interesting sources will be described and analyzed in
detail in forthcoming papers.

\section{Results}

In several objects, namely in CRL\,618, CRL\,2688, OH\,231.8+4.2,
NGC\,6302, Frosty Leo, and IRAS\,17436+5003 (HD\,161796), we can
separate the emission of the line central core (which is the dominant
peak in almost all cases) from that of the line wings, which are
expected to come, respectively, from the fossil AGB envelope and from
the fast outflows resulting from the post-AGB axial acceleration
(Sect.\ 1). We have already studied in detail the case of CRL\,618
\citep{bujetal10}; our data of CRL\,2688 and OH\,231.8+4.2, which are
also well known nebulae, will also be analyzed more in depth in
forthcoming papers.

In others (the Red Rectangle, the Boomerang Nebula, and
IRAS\,22272+5435), no line wings are detected. The Red Rectangle shows
a peculiar dynamics dominated by rotation, involving only moderate
speeds \citep{bujetal05}; see preliminary analysis of our data in
Sect.\ 3.1.  In IRAS\,22272+5435, no fast flows have been detected to
date and molecules are apparently abundant only in a relatively
slow-moving shell \citep{bujetal01,fong06}.  The Boomerang Nebula, a 
less well studied object, shows a low-velocity warm component plus,
probably, very cold high-velocity gas, which cannot be detected in
high-$J$ CO emission \citep{bujb91,sahai97}.

Finally, NGC\,7027 is a complex nebula mainly composed of a slightly
elongated (empty) shell that shows many peculiarities
\citep{naka10}. Our line profiles are composite and intense, as
expected in view of mm-wavelength data. In this source, we did not try
any simplified discussion in terms of line-core and -wing emission
because of its complex structure. The interpretation is also hindered
by the relatively large angular extent of the nebula. Our single-point
observations of CO \jdu\ (from the 30m telescope) and \jdsq\ (from
Herschel), those showing the highest angular resolution, are more
sensitive to the central part of the nebula image. Therefore, these
data tend to detect only the extreme {\em LSR} velocities ($\sim$ 10
and 40 \kms), since, due to projection effects, the emission in these
velocities comes from the central regions of the expanding shell image.
The optically thin \trece\ lines and \dieciocho\ \jsc\ show high
intensity in the central velocities; they should come from slow gas
with very high column density. \doce\ \jsc\ shows a very strong self
absorption in the blue part of the spectrum, expected in optically
thick lines from expanding gas. This self absorption is much stronger
in this line than in the \jdu\ and \juc\ ones, indicating the presence
of particularly high opacity and relatively warm gas in the
shell. Finally we note that H$_2$O lines clearly come from the fastest
gas, though in this case the spatial resolution is not very high. These
results suggest that H$_2$O formation in this C-rich environment is a
result of the shocks that accelerated the gas. Curiously, the OH emission
(observed with higher spatial resolution) is not intense in high-velocity
features.  Obviously, NGC\,7027 is one of the sources for which the
high-excitation molecular emission requires a careful modeling; results
will be soon reported \citep[][in preparation]{sant11}.

In Table 2, we show characteristic intensities of the CO lines averaged
over representative regions of the profiles, the center and the blue
and red wings, in the sources in which they can be well defined.
Intensities of H$_2$O lines in the same velocity ranges are shown in
Table 3. The adopted velocity ranges are given in those tables.  In
order to follow a procedure as objective as possible, avoiding any bias
due to {\it a priori} ideas about the nature of the sources,
intensities are averaged in all cases over velocity ranges 5 \kms\ wide
and defined following a standard procedure. This width seems a good
compromise to yield an intensity average over a relatively large number
of spectral channels, but being narrow enough to allow the separation
of each spectral feature, as we will see below, and to avoid
integrating spectral regions with too weak emission.
In the definition of these velocity ranges, we have taken into account
previous observations of low-$J$ CO lines to identify the different
components in the line profiles, which are not always well seen in our
profiles. In particular, values of the systemic and expansion
velocities of the slow components are mainly based on data and
estimates of these parameters by \cite{bujetal01}. We have also
confirmed from published maps the identification of such features in
terms of kinematical components and checked possible contaminations
between their emissions. For the line wings, the representative
velocity slots were slightly more separated from the center than the
slow-component expansion velocity, to avoid contamination from that
component, in particular the relative minima due to absorption by the
outer slow gas, which can be clearly seen in some profiles. In a few
cases, the line core may also be contaminated by emissions of fast
outflows that, due to projection effects, may show {\it LSR} velocities
close to the systemic one; we have checked for each individual nebula,
taking into account the existing information (see below), that this
contamination is negligible for the used velocity slots.  

We can see detailed information on the structure and kinematics of
CRL\,618 in \cite{sanchezc04}; maps of IRAS\,17436+5003 are shown by
\cite{ccarrizo04}. Both sources show an extended slow component, inside
which fast bipolar flows are placed, whose emissions can be well
identified in single-dish profiles \citep{bujetal01}.  The general
structure of CRL\,2688 is similar, but the fast gas structure is
complex and several collimated jets exist, see \cite{fong06} and
\cite[][note that the extended component is significantly filtered out
in these interferometric data]{cox00}. The relatively narrow velocity
slot in the center used here minimizes contamination by emission from
jets running almost in the plane of the sky, which can be present in
this source; on the other hand, the contribution of the fast component
to the single-dish profiles is well delimited by the absorption feature
seen in single-dish observations. The fast bipolar flows in Frosty Leo
are more extended than the slow molecular gas (though smaller than the
optical nebula), emissions from both components being clearly separated
in {\em LSR} velocity \citep{bujetal01,ccarrizo05}. The resulting {\em
LSR} velocity ranges taken for CRL\,618 are [$-$24.5,$-$19.5] \kms\
(line center), [$-$49,$-$44] \kms\ (blue wing) and [0,+5] \kms\ (red
wing); for CRL\,2688, we took respectively [$-$37.5,$-$32.5],
[$-$62,$-$57], and [$-$13,$-$8] \kms; for Frosty Leo we took
respectively [$-$13.5,$-$8.5], [$-$27,$-$22], and [0,+5] \kms; and for
IRAS\,17436+5003 we assumed respectively the ranges [$-$37.5,$-$32.5],
[$-$54,$-$49], and [$-$21,$-$16] \kms.

The case of OH\,231.8+4.2 is different, because it shows a very
elongated CO-rich component with a very well-defined velocity gradient
\citep{alco01}, in which the fossil AGB envelope is not clearly
identified. In this nebula, we selected the {\em LSR} ranges
[+30.5,+35.5] (line core), [0,+5] (blue wing), [+61,+66] (red wing).  
The two high-velocity ranges correspond to regions clearly different
from the central condensation \citep[in clumps I2 and I4 in the
nomenclature of][which have been very probably accelerated along the
nebular axis]{alco01}, but not very far from the center, to be sure
that their emissions are intense and well within the telescope
beamsize. The central velocity range we selected represents the central
clump, which, in view of its structure, kinematics, and chemistry, does
not seem to be affected by post-AGB shocks \citep{alco01,sanchezc97}. 

In NGC\,6302, most of the molecular gas occupies an equatorial
torus or disk, in moderate expansion ($\sim$ 15 \kms) and perpendicular
to the spectacular optical lobes, see maps in mm-wave CO by
\cite{peretto07} and \cite{trung08}. These authors argue that the
equatorial structure is probably the remnant of the (late) AGB mass
loss, not strongly altered during the post-AGB phase.  The two-horn
dominant feature of our profiles appears as a result of the shape and
relatively constant velocity of this component. There are also
high-velocity clumps extending partially along the lobes; but they are
very patchy and only those at relatively negative velocity (associated
to the southeast part of the nebula) are intense enough to be analyzed
here. Efficient photodissociation due to the UV radiation from the
very hot central star is probably the reason for the distribution of
molecule-rich gas found in NGC\,6302.  Therefore, in this source we
just considered its blue wing, range [$-$53,$-$48] \kms, and the
line core, range [$-$25,$-$20] \kms. We will assume that these
velocity ranges represent respectively the high-velocity molecular
gas and the unaccelerated remnant of the AGB shell, but keeping in mind
that the nature of both components in this source may be different than
in other nebulae studied here and the existence of large amounts of
nebular gas not detected in molecular emission. 

\begin{table*}
\caption{Characteristic intensities of the central core
  (or peak) and wings of CO lines in the sources in which they can easily
  be identified. }
\begin{center}
\begin{tabular}{llcccccc}
molecule & line & central core & blue wing & red wing & wing/core
& line-core/2--1-core & line-wing/2--1-wing \\ 
         &      & average $T_{\rm mb}$(K) & av.\ $T_{\rm mb}$(K) & av.\
$T_{\rm mb}$(K) & & & \\
\hline
\multicolumn{2}{r}{{\em LSR} velocity ranges (\kms):} & [$-$24.5,$-$19.5] &
	    [$-$49,$-$44] & [0,+5] &  &  \\
CRL\,618 & \doce\ $J$=2--1   & 13 & 0.62 & 0.92 & 0.07 & 1 & 1 \\
         & \doce\ $J$=6--5   & 1.08 & 0.35 & 0.29 & 0.3 & 0.08 & 0.42 \\
         & \doce\ $J$=10--9  & 1.8 & 1.0  & 0.74 & 0.47 & 0.14 & 1.1 \\
         & \doce\ $J$=16--15 & 2.25  & 1.5  & 0.89 & 0.53 & 0.17 & 1.55 \\
         & \trece\ $J$=2--1   & 2.3  & 0.006 & 0.24 & 0.053 & 1 & 1 \\
         & \trece\ $J$=6--5   & 0.29 & 0.04 & 0.057 & 0.17 & 0.11 & 0.4 \\
         & \trece\ $J$=10-9   & 0.34 & 0.086 & 0.085 & 0.25 & 0.15 & 0.70 \\
         & \trece\ $J$=16--15 & 0.14 & n.d. & n.d. & -- & 0.06 & -- \\
\hline
\multicolumn{2}{r}{{\em LSR} velocity ranges (\kms):} & [$-$37.5,$-$32.5] &
	    [$-$62,$-$57] & [$-$13,$-$8] &  &  \\
CRL\,2688 & \doce\ $J$=2--1   & 24.1 & 5.3 & 1.24 & 0.14 & 1 & 1 \\
          & \doce\ $J$=6--5   & 3.2 & 0.98 & 0.28 & 0.19 & 0.13 & 0.19 \\
          & \doce\ $J$=10--9  & 5.55 & 2.0 & 0.55 & 0.23 & 0.23 & 0.39 \\
          & \doce\ $J$=16--15 & 3.4 & 1.3 & 0.245 & 0.23 & 0.14 & 0.235 \\ 
          & \trece\ $J$=2--1 & 6.5 & 0.6 & 0.22 & 0.063 & 1 & 1 \\
          & \trece\ $J$=6--5 & 0.92 & 0.13 & 0.042 & 0.09 & 0.14 & 0.21 \\
          & \trece\ $J$=10-9 & 0.84 & 0.085 & 0.028 & 0.067 & 0.12 & 0.14 \\
& \trece\ $J$=16--15 & 0.26 & $\sim$ 0.02 & $\sim$ 0.04 & $\sim$ 0.11 &
0.04 & $\sim$ 0.07 \\
\hline
\multicolumn{2}{r}{{\em LSR} velocity ranges (\kms):} & [+30.5,+35.5] &
	    [0,+5] & [+61,+66] &  &  \\
OH\,231.8+4.2 & \doce\ $J$=2--1   & 3.4  & 1.3  & 1.6 & 0.43 & 1 & 1 \\
        & \doce\ $J$=6--5  & 0.33 & 0.047 & 0.12 & 0.26 & 0.1 & 0.058  \\
        & \doce\ $J$=10--9 & 0.24 & -- & --  & \lsim\ 0.2 & 0.07 &
\lsim\ 0.04 \\
  & \doce\ $J$=16-15 & \lsim\ 0.1 & --  & -- & -- & \lsim\ 0.03 & -- \\
        & \trece\ $J$=2--1 & 1.25 & 0.25 & 0.35 & 0.24 & 1 & 1 \\
        & \trece\ $J$=6--5 & 0.15 & -- & 0.03 & 0.1 & 0.12 & 0.05 \\
        & \trece\ $J$=10--9 & 0.085 & -- & -- & \lsim\ 0.3 & 0.07 &
\lsim\ 0.1 \\ 
  & \trece\ $J$=16-15 & \lsim\ 0.2 & --  & -- & -- & \lsim\ 0.2 & -- \\
\hline
\multicolumn{2}{r}{{\em LSR} velocity ranges (\kms):} & [$-$25,$-$20] &
         [$-$53,$-$48] & -- &  &  \\
NGC\,6302 & \doce\ $J$=2--1   & 4.3 & 3.7  & -- & 0.85  & 1    & 1 \\
          & \doce\ $J$=6--5   & 0.705 & 0.14 & -- & 0.2 & 0.14 & 0.038 \\
          &\doce\ $J$=10--9   & 1.29 & 0.135 & -- & 0.11 & 0.29 & 0.037 \\
          &\doce\ $J$=16--15  & 1.0 & \lsim\ 0.1 & -- & \lsim\ 0.1 & 0.23
& \lsim\ 0.02 \\
        & \trece\ $J$=2--1 & 1.3 & 1.2  & -- & 0.9  & 1  & 1 \\  
        & \trece\ $J$=6--5 & 0.255 & 0.033 & -- & 0.13 & 0.2 & 0.027 \\
         &\trece\ $J$=10--9  & 0.36 & 0.014 & -- & 0.039 & 0.27 & 0.012 \\
\hline
\multicolumn{2}{r}{{\em LSR} velocity ranges (\kms):} & [$-$13.5,$-$8.5] &
	    [$-$27,$-$22] & [0,+5] &  &  \\
Frosty Leo & \doce\ $J$=2--1 & 1.3   & 1.05  & 0.95  & 0.77  & 1 & 1 \\
           & \doce\ $J$=6--5 & 0.031 & 0.017 & 0.016 & 0.53 & 0.0235 & 0.0165 \\
           & \trece\ $J$=2--1 & 0.64 & 0.58  & 0.53  & 0.87   & 1 & 1 \\
           & \trece\ $J$=6--5 & 0.016 & 0.01 & 0.0075 & 0.55 & 0.025 & 0.016 \\
\hline
\multicolumn{2}{r}{{\em LSR} velocity ranges (\kms):} & [$-$37.5,$-$32.5] &
	    [$-$54,$-$49] & [$-$21,$-$16] &  &  \\
IRAS\,17436+5003 & \doce\ $J$=2--1 & 1.75 & 0.08  & 0.09  & 0.05  & 1 & 1 \\
(HD\,161796) & \doce\ $J$=6--5 & 0.94 & 0.01 & 0.01  & 0.11  & 0.054 & 0.12 \\
           & \doce\ $J$=10--9 & 0.06 & 0.022 & 0.016 & 0.32 & 0.034 & 0.225 \\ 
           & \trece\ $J$=2--1 & 0.72 & 0.08  & 0.09  & 0.12 & 1 & 1 \\
           & \trece\ $J$=6--5 & 0.033 & 0.016 & 0.0105 & 0.4 & 0.046 & 0.155 \\
\hline
\end{tabular}
\end{center}
{\vspace{-.2cm}\bf Notes:} See discussion on the respective velocity ranges in
  the text. Ratios wing/core and with respect to the \jdu\ lines (from
  30-m telescope data) are also indicated; in these cases, the average
  of both wings is taken. 
\end{table*}

\begin{table*}
\caption{Characteristic intensities of the central core (or peak) and
  wings of H$_2$O and OH lines in the cases in which they can be
  measured.}
\begin{center}
\begin{tabular}{llccc}
molecule & line & central core & blue wing & red wing \\ 
         &      & average $T_{\rm mb}$(K) & av.\ $T_{\rm mb}$(K) & av.\
$T_{\rm mb}$(K) \\
\hline
\multicolumn{2}{r}{{\em LSR} velocity ranges (\kms):} & [$-$24.5,$-$19.5] &
	    [$-$49,$-$44] & [0,+5] \\
CRL\,618 & H$_2$O  1$_{1,0}$--1$_{0,1}$ & 0.15 & 0.035 & ISM?  \\
         & H$_2$O  1$_{1,1}$--0$_{0,0}$ & 0.34 & 0.095 & ISM?  \\
         & H$_2$O  3$_{2,1}$--3$_{1,2}$ & $\sim$0.2 & -- & -- \\
         & OH $^2\Pi_{1/2}$ 3/2$^-$--1/2$^+$ & 0.28 & 0.07 & -- \\
\hline
\multicolumn{2}{r}{{\em LSR} velocity ranges (\kms):} & [$-$37.5,$-$32.5] &
	    [$-$62,$-$57] & [$-$13,$-$8]   \\
CRL\,2688 & H$_2$O  1$_{1,1}$--0$_{0,0}$   & 0.16 & 0.05 & 0.02  \\
\hline
\multicolumn{2}{r}{{\em LSR} velocity ranges (\kms):} & [+30.5,+35.5] &
	    [0,+5] & [+61,+66]  \\
OH\,231.8+4.2 & H$_2$O  1$_{1,0}$--1$_{0,1}$ & 0.29  & 0.026 & 0.078 \\
  & H$_2$O  1$_{1,1}$--0$_{0,0}$ & 0.55 & -- & 0.16  \\
  & H$_2$O  3$_{2,1}$--3$_{1,2}$ & 1.6 & 0.05 & 0.41 \\
  & H$_2$O  3$_{1,2}$--2$_{2,1}$ & 0.94 & -- & 0.14  \\
  & H$_2$O  4$_{2,2}$--4$_{1,3}$ & 1.26 & 0.03  & 0.03 \\
  & H$_2$O  6$_{3,3}$--6$_{2,4}$ & 0.17 & -- & -- \\
  & H$_2$O  7$_{3,4}$--7$_{2,5}$ & 0.13 & -- & -- \\
  & OH $^2\Pi_{1/2}$ 3/2$^-$--1/2$^+$ & 1.04 & 0.083 & 0.38 \\
\hline
\multicolumn{2}{r}{{\em LSR} velocity ranges (\kms):} & [$-$25,$-$20] &
         [$-$53,$-$48] & -- \\
NGC\,6302 & H$_2$O 1$_{1,0}$--1$_{0,1}$ & 0.055 & 0.013 & -- \\
          & H$_2$O 1$_{1,1}$--0$_{0,0}$ & 0.085 & -- & -- \\
  & OH $^2\Pi_{1/2}$ 3/2$^-$--1/2$^+$ & 0.25 & 0.02 & -- \\
\hline
\multicolumn{2}{r}{{\em LSR} velocity ranges (\kms):} & [$-$13.5,$-$8.5] &
	    [$-$27,$-$22] & [0,+5]   \\
Frosty Leo &  H$_2$O  1$_{1,0}$--1$_{0,1}$ & 0.025 & 0.019 &0.024 \\
\hline
\end{tabular}
\end{center}
{\vspace{-.2cm}\bf Notes:} See discussion on the respective velocity
  ranges in the text.  ISM?: no value is given because of possible
  interstelar contamination.
\end{table*}

We have computed wing/core CO intensity ratios and ratios between the
observed lines and previous data of the CO $J$=2--1 transitions
\citep[for previous observations, see Table 1,][and references
therein]{bujetal01}. For the wings, we have taken the average of both
line wings, except for NGC\,6302, in which only the blue wing is
considered.  See results in Table 2 and in Figs. 15, 16, 17.  It is
obvious from these ratios that there is a wide variety of excitation
conditions in the high-velocity (presumably shocked) gas. In Fig.\ 15
we can see the wing/core intensity ratios for three well known
sources. These variations are not due to significant differences in the
central core intensities, for which the ratio with respect to the 2--1
transition varies very slightly from line to line and from source to
source.  The \doce\ and \trece\ line-wing ratios with respect to the
\jdu\ line are shown in Fig.\ 16, these ratios are also a measurement
of the excitation state of the high-velocity gas. The fast increase of
the wing/core ratio with the energy of the levels in CRL\,618 shows
that the temperature in its fast outflow must be high; the case of
OH\,231.8+4.2 is completely different, with a clear decrease of this
ratio, showing a cool fast flow; CRL\,2688 seems to be an intermediate
case.  For CRL\,618 and OH231.8+4.2 we have an accurate idea of the
temperature in the outflows from detailed model fitting of Herschel CO
data, which indeed confirms this preliminary interpretation. Bujarrabal
et al.\ (2010) showed that the temperature of the fast flow of CRL\,618
is high, $T_{\rm k}$ \gsim\ 200 K. However, we know that the fast flow
temperature in OH\,231.8+4.2 is always cooler than about 30 K
\citep[][in preparation]{alco11}.


\begin{figure}
\vspace{-0.cm}
{\hspace{-0cm}\vspace{-0.cm}\rotatebox{270}{\resizebox{6.1cm}{!}{ 
\includegraphics{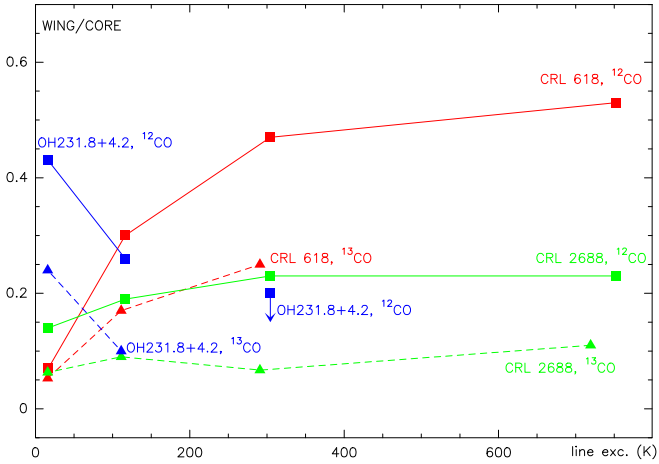}
}}}
\caption{Line-wing over line-core main-beam temperature ratio of the
observed CO lines in the well studied sources CRL\,618, CRL\,2688, and
OH\,231.8+4.2 ($J$=2--1 data are taken from previous works); see Table
2. Note the different behavior of the relative line-wing intensity
between CRL\,618, which shows particularly hot high-velocity outflows,
and OH\,231.8+4.2, in which the flows remain quite cold. CRL\,2688 is
an intermediate case. Squares (continuous line) represent $^{12}$CO
data and triangles (dashed lines) show $^{13}$CO data. Each rotational
transition is represented by the energy (in K) of its upper level. }
\end{figure}

\begin{figure}
\vspace{-0.0cm}
{\hspace{-.cm}\vspace{-0.cm}\rotatebox{270}{\resizebox{6.1cm}{!}{ 
\includegraphics{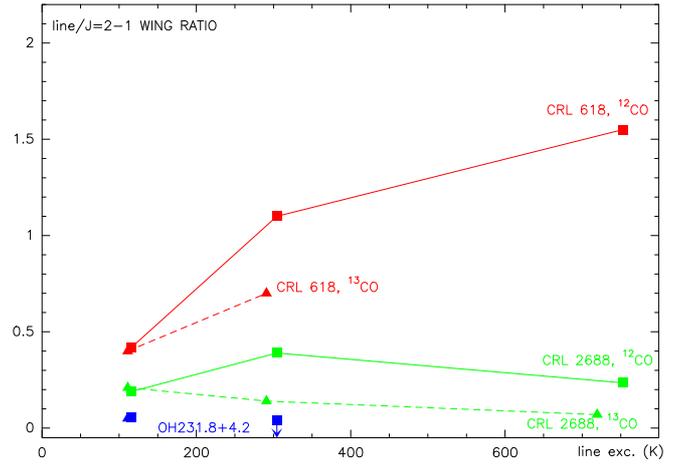}
}}}
\caption{Wing intensity ratio between the observed lines and the CO
$J$=2--1 transition (from previous works) in the well studied sources
CRL\,618, CRL\,2688, and OH\,231.8+4.2; see Table 2. A comparison
between these ratios and estimates from our simple, general models
reveals values of the characteristic temperature compatible with
estimates from detailed models (see text). Squares (continuous line)
represent $^{12}$CO data and triangles (dashed lines) show $^{13}$CO
data. Each rotational transition is represented by the energy (in K) of
its upper level. }
\end{figure}

\begin{figure}
\vspace{-0.0cm}
{\hspace{-.0cm}\vspace{-.0cm}\rotatebox{270}{\resizebox{6.1cm}{!}{ 
\includegraphics{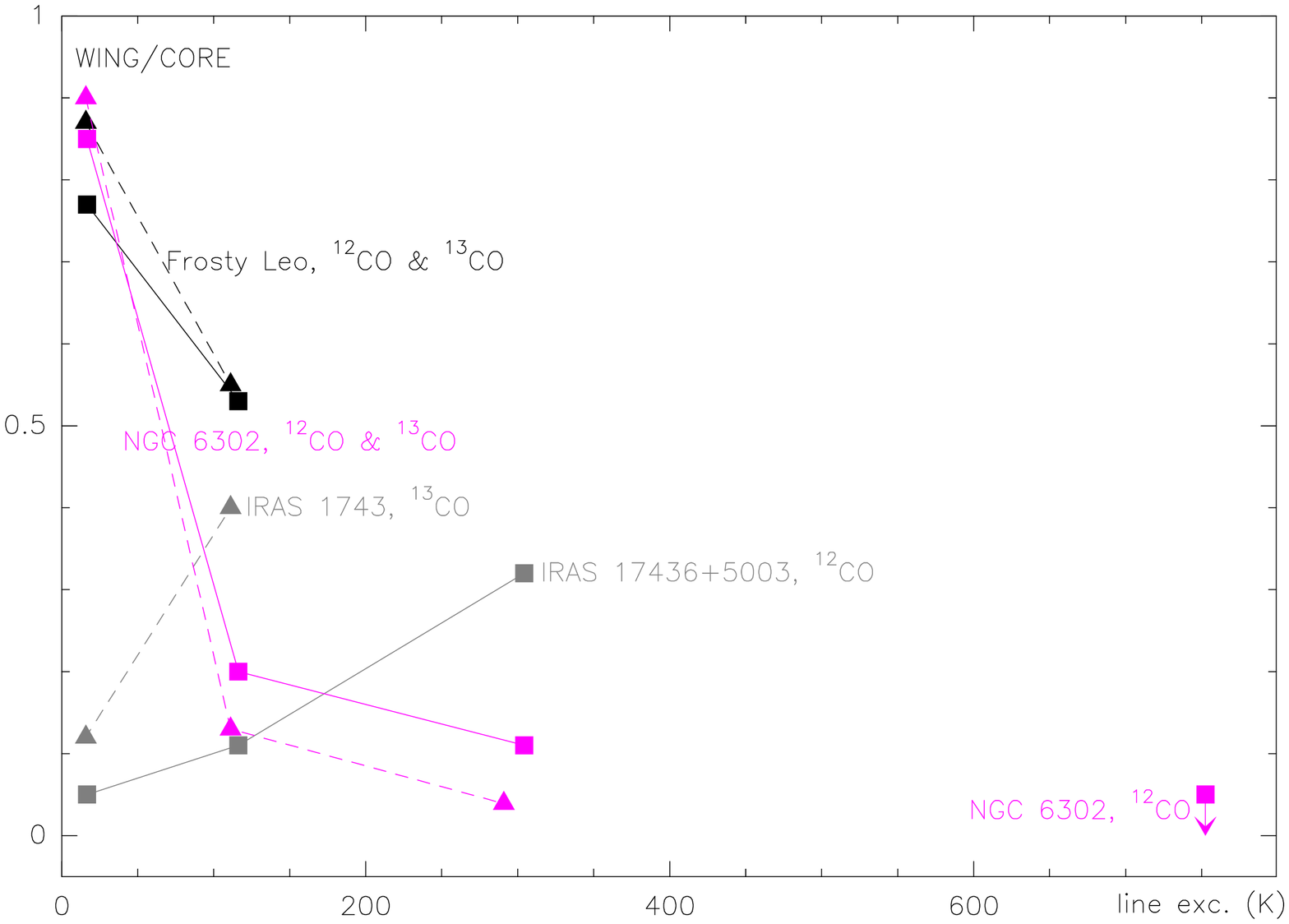}
}}}
\caption{Line-wing over line-core main-beam temperature ratio of the
observed CO lines in sources that are less well studied in molecular
emission: Frosty Leo, NGC\,6302, and IRAS\,17436+5003. Also in these
cases, we can identify hotter and cooler fast outflows. Squares
(continuous line) represent $^{12}$CO data and triangles (dashed lines)
show $^{13}$CO data. Each rotational transition is represented by the
energy (in K) of its upper level.  }
\end{figure}

\subsection{Comparison with simplified calculations of CO line emission}

Our line profiles must be compared with theoretical predictions in
order to derive the physical properties of the nebular gas from the
empirical data. Here we just present a simplified description
of the analysis to be performed, mostly based on single-component,
standard calculations. More detailed modeling of the nebulae will be
performed in dedicated papers, where their complex structures will be
taken into account.

A comparison of the wing intensity ratio between our lines and $J$=2--1
(see Fig.\ 16) with theoretical predictions can be used to perform a
rough estimate of the physical conditions in the fast outflows in our
sources, particularly the characteristic temperature. Results from
plain rotational diagrams, which require optically thin and thermalized
lines, are not accurate in our cases, since lines a high as $J$=16--15
are only thermalized for densities \gsim\ 10$^7$ cm$^{-3}$ and
optically thin emission is not expected in \doce\ lines (even \trece\
lines seem often to present moderate values of the optical depth).

To perform a very simplified estimate of the excitation conditions in
our nebulae, we have carried out theoretical calculations of line
ratios for a variety of values of the density and temperature ($n_{\rm
tot}$ and $T_{\rm k})$, using the well known {\it LVG} (large velocity
gradient) formalism. We have assumed standard values for the rest of
the parameters that enter the calculation: distance to the center $R$ =
6 10$^{16}$ cm, expansion velocity $V_{\rm exp}$ = 30 \kms, and
relative abundances $X$($^{12}$CO) = 2 10$^{-4}$, $X$($^{13}$CO )=
10$^{-5}$. It is obvious that those conditions cannot be applied to all
nebulae, since we assume a single value for the product $R\,X/V_{\rm
exp}$ ($n_{\rm tot}R\,X/V_{\rm exp}$ represents the column density in
the opacity under the {\it LVG} approximation). However, this approach
is enough for our purpose of understanding the general properties of
the line intensities from the theoretical point of view, particularly
to give examples of the behavior of the line intensities and ratios
depending on the temperature and density. Opacity effects are taken
into account in the calculations, under the usual assumptions of this
formalism.

\begin{figure}
{\hspace{1.0cm}\vspace{-0.cm}\rotatebox{270}{\resizebox{8cm}{!}{ 
\includegraphics{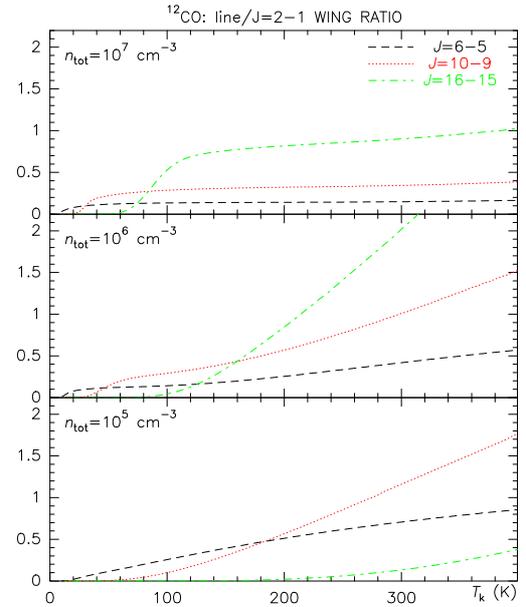}
}}}
\caption{Predictions for the wing intensity ratios between the observed
  \doce\ lines and the \doce\ $J$=2--1 transition obtained from our
  non-LTE calculations for our three lines, \jsc, \jdn, and \jdsq,
  respectively dashed (black in electronic version), dotted (red), and
  dash-dot (green) lines. Intensity ratios are shown as functions of
  the kinetic temperature in the three panels for, respectively, 
  densities equal to 10$^7$, 10$^6$, and 10$^5$ cm$^{-3}$. }
\end{figure}

\begin{figure}
{\hspace{1.0cm}\vspace{-0.cm}\rotatebox{270}{\resizebox{8cm}{!}{ 
\includegraphics{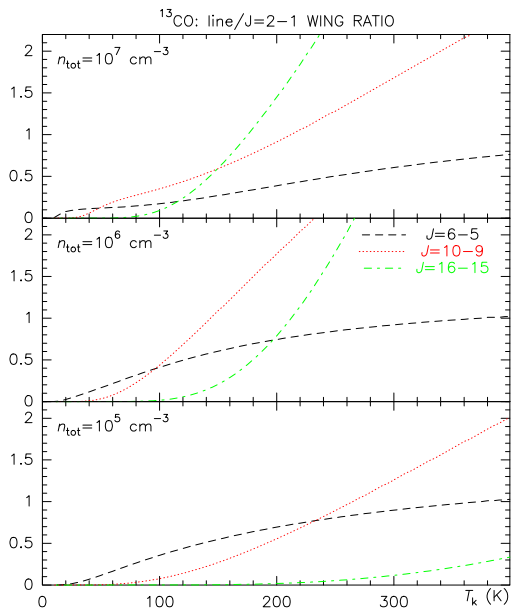}
}}}
\caption{Predictions for the wing intensity ratios between the observed
  \doce\ lines and the \trece\ $J$=2--1 transition obtained from our
  non-LTE calculations for our three lines, \jsc, \jdn, and \jdsq,
  respectively dashed (black in electronic version), dotted (red), and
  dash-dot (green) lines. Intensity ratios are shown as functions of
  the kinetic temperature in the three panels for, respectively, 
  densities equal to 10$^7$, 10$^6$, and 10$^5$ cm$^{-3}$. }
\end{figure}

The {\em LVG} approximation itself is fully justified in our case,
since it is well known that PPNe often show strong velocity gradients,
with radial expansion velocities increasing more or less proportionally
to the distance to the star. For this reason, we do not expect strong
radiative interactions between points separated by a long distance and
showing very different physical conditions. This approximation is
appropriate for our purposes also because it does not require that
details on the large-scale nebula structure are included in the code,
provided that no radiative interaction between distant points takes
place, allowing a general description of the line excitation for a
number of different objects. Moreover, in our bipolar flows, often
elongated and showing large velocity gradients, the emission at each
{\it LSR} velocity must come from a relatively compact region. In this
way, the code gives characteristic values of the brightness for each
line, from which we can readily calculate line intensity ratios.  Line
intensities at a given velocity can be also estimated if we assume an
angular size for the emitting region.

Results of our calculations for \trece\ and \doce\ are shown in Figs.\
18, 19, 20, and 21. In our analysis, we will focus on the line
intensity ratios, since ratios depend only slightly on the linear and
angular extent of the source and are more easily described by
general models.  Figs.\ 18 and 19 show predicted line ratios for
several densities and temperatures, within the ranges needed to explain
our data. In this case we have assumed that the lines come from a
region significantly smaller than the telescope spatial resolution
(which is justified in our case), so that the comparison of the
predicted line brightness ratios must be corrected for the respective
telescope beam widths (half-maximum beam widths of the Herschel
observations are given in Sect.\ 2; we have adopted a value of 12$''$
for the 30m-telescope observations of CO $J$=2--1). In Figs.\ 20 and
21, we show predicted intensities; normalized units are used to allow a
better comparison between the intensities of the different lines and
sets of physical conditions.  The chosen units correspond to $T_{\rm
mb}$(K) calculated assuming that the model nebula with density $n_{\rm
tot}$ = 10$^6$ cm$^{-3}$ occupies a region 1$''$ wide, and taking the
different telescope resolutions into account (which are again much
larger than the emitting region). To make easier the comparison between
predictions for other densities, we have taken varying values of the
emitting angular size, inversely proportional to the density, such that
the total emitting mass is similar for the different model
condensations. We can check that the intensities are of the order of
those observed: for instance a model clump $\sim$ 2$''$ wide, with
$T_{\rm k}$ = 100 K and $n_{\rm tot}$ = 10$^6$ cm$^{-3}$ would yield a
typical brightness $T_{\rm mb}$(\doce\ \jdn) $\sim$ 1 K, similar to
some of our observed line-wing intensities.


The dependence of the predicted high-$J$ line intensity on the
temperature is clear, as expected, and can be seen both in line ratios
(Figs.\ 18 and 19) and intensities (Figs.\ 20 and 21). The intensity of
the relevant lines increases with temperature, but from a certain value
of $T_{\rm k}$ it saturates and even decreases. Of course, the
temperature limit of intensity saturation increases for higher values
of $J$, which imply higher level energies. We can also see that the
line intensity ratios tend to increase with the volume density,
particularly for the highest $J$-values, but the dependence is less
clear than with respect to the temperature. The general properties of
the excitation of the high-$J$ CO lines discussed here are therefore
different than those of low-excitation transitions of high dipole
moment species, like for instance HCO$^+$, SiO, etc, often observed
with ground-based telescopes. In these low-energy transitions with
large radiative decay rates, the emission depends strongly on the
density (the intensity per unit volume being roughly proportional to
$n_{\rm tot}^2$), because collisional de-excitation is much less
probable than radiative transitions. However, its dependence on the
temperature is in fact very moderate, because the intensity saturation
level is low.

From comparison of our \trece\ observational results with model
predictions in Fig.\ 19, we derive that the high-velocity CO emission
in CRL\,618 essentially comes from a region at $T_{\rm k}$ $\sim$ 150 K
and with a density $n_{\rm tot}$ $\sim$ 10$^6$ cm$^{-3}$ (as we
will see below, our \doce\ data indicate that the temperature could be
somewhat higher). In OH\,231.8+4.2, our data are compatible with a
temperature $\sim$ 30 K and a lower density. Both results are in good
agreement with detailed model estimates \citep[see above discussion
and][]{bujetal10,alco11}.

Our \doce\ observations of CRL\,618 are compatible with the predictions
in Fig.\ 18 for a density of about 10$^6$ cm$^{-3}$ and a temperature
of about 270 K, higher than the value deduced from \trece\ data.  This
discrepancy may come from the fact that, in sources showing significant
variations of the density within the region emitting at a given
velocity, the optically thick \doce\ emission selects very dense
regions less efficiently than the \trece\ lines, those denser regions
being probably cooler (see discussion below on the line ratios
predicted by two-component models). \cite{bujetal10} deduce from their
detailed model fitting that the temperature of the high-velocity
outflow must vary within this range, with typical values around 200 K,
probably higher in the base of the outflow.  \cite{sanchezc04} deduce a
density similar to that given here \citep[which was also assumed
by][]{bujetal10}.  For OH\,231.8+4.2, we derive from the \doce\ lines
$T_{\rm k}$ $\sim$ 30 K and $n_{\rm tot}$ $\sim$ 10$^5$ cm$^{-3}$,
which are again comparable to the values obtained from detailed models
by \cite{alco11}.

In the case of CRL\,2688, \doce\ line ratios are compatible with a
temperature of about 100 K and $n_{\rm tot}$ $\sim$ 10$^6$
cm$^{-3}$. Such conditions are also roughly compatible with our \trece\
data. It is probable that also in this case the complex nebula harbors
gas with variable physical conditions.

As we see, even if the procedure depicted here is simple, the values we
derive are quite self-consistent and consistent with other more
accurate results from detailed modeling. 

We can also perform the same kind of analysis for the less well studied
sources NGC\,6302, Frosty Leo, and IRAS\,17436+5003. Obviously, fast
winds in NGC\,6302 and Frosty Leo show a very low excitation. In Frosty
Leo, \cite{ccarrizo05} predicted a low temperature, from low-$J$
observations, but our data indicate the presence of a slightly warmer
component at high velocity, with a temperature $\sim$ 60 K (assuming
$n_{\rm tot}$ $\sim$ 10$^4$ cm$^{-3}$, following Castro-Carrizo et
al.). In NGC\,6302, the fast gas must present a low temperature $\sim$
40 K, and the density must be of about a few 10$^5$ cm$^{-3}$; both
results are compatible with estimates by \cite{trung08} from
interferometric $J$=2--1 mapping.  The fast flow in the less well
studied IRAS\,17436+5003 must be more excited, with $T_{\rm k}$ $\sim$
60 K and $n_{\rm tot}$ $\sim$ 10$^6$ cm$^{-3}$.  In this object, the
properties of the molecular component and the evolutionary stage seem
to be similar to those of CRL\,2688 \citep[see e.g.][note that lines in
IRAS\,17436+5003 are weaker, mainly because it is more
distant]{bujetal01}.

As mentioned, these estimates are in any case crude, particularly in
view of the probably complex structure of the emitting regions. In
order to investigate those effects in general terms, we must consider
our calculations of intensity predictions in Figs.\ 20 and 21.  Let's
take a two-component case.  In general, we can see that when there is a
massive component with $n_{\rm tot}$ \lsim\ 10$^5$ cm$^{-3}$ and
$T_{\rm k}$ \lsim\ 200 K the \jdu\ emission is always dominant in it
and, even if a denser/warmer component exists, the total intensity
ratios \jdn/\jdu\ and \jdsq/\jdu\ can hardly be larger than $\sim$ 1/2.
If there is a significant component with $T_{\rm k}$ \gsim\ 200 K and
$n_{\rm tot}$ \gsim\ 10$^5$ cm$^{-3}$, its \jdn\ and \jdsq\ emission
would always be dominant over that of a diffuse/cool component. This
would explain the general trend of relatively low \jdn/\jdu\ and
\jdsq/\jdu\ ratios found in our data.

\begin{figure}
{\hspace{-.0cm}\vspace{-0.cm}\rotatebox{270}{\resizebox{6.5cm}{!}{ 
\includegraphics{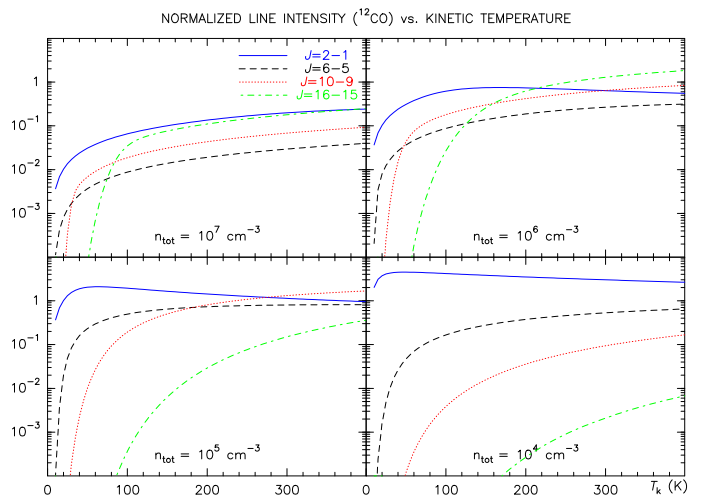}
}}}
\caption{\doce\ line intensities from our simplified model
  calculations, for several values of the temperature and density, see 
 intensity normalization procedure in Sect.\ 3.1.
}
\end{figure}

\begin{figure} 
 {\hspace{-.0cm}\vspace{-0.cm}\rotatebox{270}{\resizebox{6.5cm}{!}{ 
\includegraphics{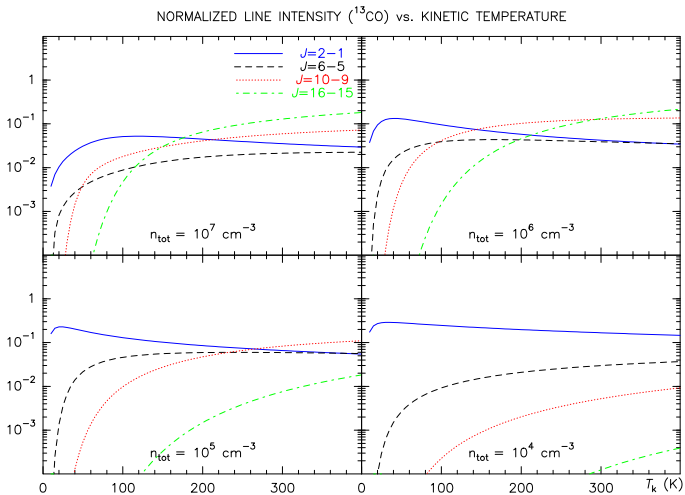}
 }}}
\caption{\trece\ line intensities from our simplified model
  calculations, for several values of the temperature and density, see 
 intensity normalization procedure in Sect.\ 3.1.
}
\end{figure}

We can also suppose that relatively diffuse, accelerated components are
systematically warmer, if the cooling after shock passage is radiative
(and, therefore, denser gas cools faster). This would yield regions
with $n_{\rm tot}$ $\sim$ 10$^5$ cm$^{-3}$ and $T_{\rm k}$ \gsim\ 200
K, whose emission is intense in \doce\ lines, but relatively weak in
\trece\ lines (with respect to dense regions). After taking the
emission of more cool/dense components into account, one would get
relatively high \doce\ \jdn/\jdu\ and \jdsq/\jdu\ line ratios, higher
than the corresponding \trece\ ratios.  Therefore, the presence of
relatively weak line-wing emission in \trece\ high-$J$ lines, often
found in our observational data, could be due to the presence of an
important relatively diffuse component, detectable in \doce\ emission
but practically not in \trece\ lines.

The presence of selfabsorption by outer, less excited layers (not
included in our simple discussion) could also help to explain this
trend. In this case, self absorption would affect mostly low-$J$ \doce\
lines, since they present the highest opacities in low-excitation
regions. This would yield relatively low \doce\ \jdu\ intensities and
then relatively high \doce\ \jdn/\jdu\ and \jdsq/\jdu\ ratios. The
existence of selfabsorption under these conditions is confirmed
by the observed absorption features in mm-wave CO profiles.

As mentioned, we have studied the case of CRL\,618 by means of detailed
modeling of its line emission, based in our good knowledge of the
structure of the nebulae from low-$J$ CO data \citep{bujetal10}, and we
are analyzing in depth the high-$J$ lines of OH\,231.8+4.2,
CRL\,2688, and NGC\,7027, which are also well known.

We are also studying in detail the CO emission from the Red Rectangle,
so here we just present a brief, simplified discussion. The Red
Rectangle is the only PPNe or young PNe in which an equatorial disk in
rotation around the central star has been well detected \citep[from
mm-wave mapping,][]{bujetal05}. This disk is, as far as we know, the
only molecule-rich component of the nebula. The structure of the line
profiles in this source is accordingly peculiar, because the outer part
of the profiles come from the inner disk regions in fast rotation
(rotation is exactly Keplerian in them), which are relatively hot and
very small. Meanwhile, the central part of the profiles mainly comes
from the outer disk. Let us take representative {\em LSR} velocities
around the systemic one $V_\pm$ = $\pm$ 1.65 \kms\ (the {\em LSR}
velocity of the center of gravity of this source is $\sim$ 0
\kms). These velocities correspond to the rotation velocity at the
point at which the dynamics change from pure Keplerian rotation to
rotation plus expansion \citep[$R_{\rm kep}$ in the notation used
by][]{bujetal05}, and there is a clear maximum of the CO lines in them.
We can see in our spectra that \doce\ \jsc\ and \jdn\ show very similar
values of the main-beam brightness at $V_\pm$. This is true even for
\jdsq, in spite of the poor-quality profile at this frequency. \trece\
\jsc\ and \jdn\ also show practically the same intensities at
$V_\pm$. As we have discussed (Sect.\ 1), mm-wave CO data are adequate
to measure densities and total masses, but often yield underestimates
of the temperature because those lines tend to select the coolest
components. So, if we assume the density estimated in these regions by
\cite{bujetal05}, 10$^5$--10$^6$ cm$^{-3}$, we conclude from comparison
with our predicted intensities, that the temperature in this
representative region is \gsim\ 150 K, at least twice as high as the
temperature deduced from mm-wave lines. This result has important
consequences for the structure and dynamics of the equatorial
disk in the Red Rectangle. Under the framework of the standard theory
of the passive rotating disks, the gas temperature directly controls
the disk extent from the equator (but comparison with observations is
difficult, because the angular resolution of the existing maps is not
high enough to properly measure the disk width). The presence of these
high temperatures is also the most probable explanation of the
detection of evaporation of gas in this source (i.e.\ of expansion from
a certain radius $R_{\rm kep}$); in fact, we can readily show that the
thermal velocity dispersion becomes comparable to the rotation velocity
at $R_{\rm kep}$ for $T_{\rm k}$ $\sim$ 200 K.

\subsection{Comparison with previous estimates}

As mentioned in Sect.\ 1, warm molecular gas in CRL\,618 and
CRL\,2688 was studied in a few previous papers. 

\cite{justtanont00} deduce the presence of warm gas from ISO data of
\doce\ emission, for transitions between $J$=14--13 and
$J$=37--36. These authors clearly detect the presence of gas at
temperatures higher than the typical values found here, of $\sim$ 380 K
in CRL\,2688 and $\sim$ 700 K in CRL\,618, using simple rotational
diagrams. This result is not unexpected, because higher-$J$ transitions
tend to come from hotter regions.  However, the lack of information on
the profile shapes (ISO spectra did not resolve the line profiles)
prevents any attempt to discern from which nebular components such
emissions arise. In the case of CRL\,618, this hot component could be
very close to the inner HII region, but any conclusion on its
properties remains very uncertain. We finally point out that it is well
known that simple rotational diagrams can overestimate the
characteristic rotational temperature, since lower transitions
typically show higher opacities.

\cite{mpb92} observed cm-wave inversion transitions of NH$_3$ up to
$J,K$=7,7, requiring several hundred K to be excited. They measured a
typical temperature of about 150 K in CRL\,2688, compatible with our
results, but without a clear distinction between the nebular
components. In CRL\,618, they detect lines mostly in absorption against
the central free-free continuum source. They were able to identify two
independent components. One of them, in slow expansion and showing a
temperature of about 270 K, is very probably the slow-expanding inner
shell responsible for the narrow line core found from Herschel high-$J$
CO data, see \cite{bujetal10}. The other component corresponds to the
fast outflow also discussed by \cite{bujetal10}, and its temperature
from NH$_3$ data is about 150 K. The temperature derived by
\cite{mpb92} for the slow component is very similar to that of the very
inner central region in slow expansion studied by means of our CO data;
but the NH$_3$ temperature of the fast outflow is slightly lower than
the one found from our results. We note that the fact that
\cite{mpb92} detect lines in absorption makes it very difficult to
identify which part of the fast outflow is actually absorbing the
continuum, perhaps the warmest regions of the outflow are not placed in
front of the very small continuum source and were not actually detected
in ammonia lines.

Finally, we mention the observations by \cite{pardo04} of
CRL\,618 in a high number of rotational transitions of HC$_3$N in the
ground and excited vibrational states. These authors deduce the
presence of a central, molecule-rich region with $T_{\rm k}$ $\sim$
250--300 K. The excitation conditions and extent of this component are
very similar to those deduced by \cite{bujetal10} for the inner nebula
from the Herschel CO profiles.

\section{Conclusions}

We present observations of FIR/sub-mm molecular lines in ten
protoplanetary or young planetary nebulae (PPNe, PNe), see Table 1,
obtained with the heterodyne spectrometer HIFI, on board the Herschel
Space Telescope. We focused on observing CO and H$_2$O rotational
lines. We have detected \doce\ and \trece\ \jsc, \jdn, and \jdsq\
lines, and we also report observations of H$_2$O 1$_{1,0}$--1$_{0,1}$,
1$_{1,1}$--0$_{0,0}$, 3$_{2,1}$--3$_{1,2}$, 3$_{1,2}$--2$_{2,1}$,
4$_{2,2}$--4$_{1,3}$, 6$_{3,5}$--6$_{2,4}$, and 7$_{3,4}$--7$_{2,5}$
lines. \doce\ was detected in all the sources. At least one H$_2$O line
was detected from all but three objects, IRAS\,22227+5435, the Red
Rectangle, and IRAS\,17436+5003. In the most intense emitters, we have
also found other lines within the (wide) receiver bands; the NH$_3$
$J,K$=1,0--0,0 and OH $^2\Pi_{1/2}$ 3/2--1/2 rotational transitions are
particularly interesting; we also detected lines of C$^{18}$O,
H$_2$$^{18}$O, vibrationally excited H$_2$O, SiO, HCN, and CN.

The obtained full-band spectra are presented in the Appendix (in
electronic form only). Selected lines are shown in Figs.\ 1 to 14.

Particularly rich are the spectra of CRL\,618, CRL\,2688, NGC\,7027,
and OH\,231.8+4.2, which have been or will soon be discussed in
dedicated papers. Several spectra of CRL\,618 were presented in a
preliminary paper \citep{bujetal10}; here we show again those spectra,
after careful reanalysis and calibration, as well as new recent
observations. 

We have performed a simplified interpretation of the CO line intensity
in terms of general excitation state of the two main components of the
profiles: the line core and the line wings. It is known (Sects.\ 1, 3)
that the line core comes from a slowly expanding component that can be
identified with the remnant of the AGB shell. On the other hand, the
line wings come from fast bipolar outflows, which are thought to be
accelerated and heated by shocks during the post-AGB phase. We argue
(Sects.\ 1 and 3.1) that those FIR/submm lines are the best tool to
study intermediate-excitation regions, in the temperature ranges 100 K
\lsim\ $T_{\rm k}$ \lsim\ 1000 K, which are not well probed by mm-wave
or optical/NIR observations.

In Table 2, we give typical CO intensities of these features, as well as
intensity ratios between them or between the observed lines and the
\jdu\ line (from IRAM 30-m telescope observations), both for \doce\ and
\trece. Significant ratios are represented in Figs.\ 15 to 17, see full
description of the parameters in Sect.\ 3. As we can see, in CRL\,618
the line wings are particularly intense and relatively more intense
for the highest-$J$ transitions, indicating particularly high
temperatures in the fast gas in this source, as already found 
by \cite{bujetal10}. In others, notably in OH\,231.8+4.2 and in
agreement with results by \cite{alco11}, the fast gas seems
significantly cooler. 

We also present theoretical predictions of FIR/submm line intensities
for wide ranges of the kinetic temperature and density, performed for a
quite general case, see details in Sect.\ 3.1 and results in Figs.\ 18
to 21. From comparison between these theoretical expectations and our
data, we have estimated characteristic physical conditions of the fast
gas in six nebulae, focusing on the kinetic temperatures.

We must remember that such a comparison is not straightforward, in
particular because of the very probable complex structure and dynamics of
the emitting regions. In fact, in view of some general trends in the
comparison of our observational and theoretical results, we argue
(Sect.\ 3.1) that several components may be systematically present and
that the densest parts of the fast gas may be systematically cooler (see
below). 

\begin{table}
\caption{Characteristic temperatures of the fast outflows derived in
  this work, compared
  with other nebular parameters. }
\begin{center}
\begin{tabular}{l|cccc}
source & $T_{\rm k}$ & flow lifetime & $n_{\rm tot}$ & refer.\ \\ 
\hline
CRL\,618    & 200 K & \lsim\ 100 yr & 5 10$^6$ cm$^{-3}$  &  $2$ \\
CRL\,2688   & 100 K & 200 yr  &  $\sim$ 10$^6$ cm$^{-3}$  & ${1,3}$ \\ 
OH\,231.8+4.2 & 30 K & 1000 yr  & $\sim$ 2 10$^5$ cm$^{-3}$ & $4$ \\ 
NGC\,6302   & 40 K & \lsim \ 800 yr & 2 10$^4$ cm$^{-3}$ & $5$ \\
Frosty Leo  & 60 K & $\sim$ 1700 yr & 10$^4$ cm$^{-3}$ & $6$ \\
IRAS\,17436+5003 & 60 K & $\sim$ 500 yr  & $\sim$ 10$^6$ cm$^{-3}$ & $1,7$ \\
\hline
\end{tabular}
\end{center}
{\vspace{-.2cm}\bf References: $1$:} this paper (Sect.\
  3.1); $2$: \cite{sanchezc04}; $3$: \cite{cox00}; $4$:
  \cite{alco01}; $5$: from data in \cite{trung08}; $6$:
  \cite{ccarrizo05}; 7: \cite{hoogzaad02,ccarrizo04}.
\end{table}

In the best studied objects, information obtained from previous
investigations help us to understand the contributions of the various
nebular components, and our discussion is more reliable. For CRL\,618
and OH\,231.8+4.2, we can compare our estimates with results from
detailed modeling of the FIR/sub-mm line emission and we find that both
results are quite consistent. Therefore, we believe that our estimates
of the density and temperature, although simple, are good
approximations to the average physical conditions in the fast outflows
of the studied nebulae. We derive a relatively high temperature for the
high-velocity gas in CRL\,618, with typical values $T_{\rm k}$ $\sim$
200 K, probably higher in the base of its (relatively compact) bipolar
outflow. However the fast gas in OH\,231.8+4.2 is quite cold, at around
30 K, and shows no signs of significantly higher temperatures in any
part of its very elongated and extended outflow. In CRL\,2688, a nebula
that is also relatively well studied, the characteristic temperature of
the high-velocity gas is intermediate, $T_{\rm k}$ $\sim$ 100 K. Table
4 shows a summary of the derived characteristic temperatures, as well
as the time elapsed since the acceleration of bipolar flows (or flow
lifetime) and an estimate of the characteristic density in that
component. (In the estimates of the lifetimes and densities, we take
results from previous studies into account, see references in the
Table.) As we can see, there is a clear relation between the
temperature of the fast gas and the time elapsed since its
acceleration, with clear extreme values for OH\,231.8+4.2 and CRL\,618,
and intermediate values for CRL\,2688 and other nebulae. We interpret
this result as showing that the gas was accelerated and heated by the
passage of a shock and it is being cooled down since then.  The density
seems also to play a role, as expected, since radiative cooling
increases with the gas density. For instance, the relatively diffuse
gas in Frosty Leo and NGC\,6302 is somewhat warmer than in
OH\,231.8+4.2, for comparable times.

We have also discussed briefly other cases of interest: NGC\,7027 and
the Red Rectangle (Sect.\ 3). NGC\,7027 shows complex profiles
corresponding to its well known very complex structure
\citep{naka10}. The molecule-rich shell is elongated and very
inhomogeneous and shows a composite velocity field. Detailed modeling
is absolutely necessary to understand the molecular emission from this
nebula. The molecular gas in the Red Rectangle forms an equatorial disk
in rotation \citep[][it is the only PPNe or PNe in which gas in
rotation has been clearly detected]{bujetal05}. From a first-approach
comparison between our profiles in the Red Rectangle and theoretical
predictions, we derive a characteristic temperature $T_{\rm k}$ \gsim\
150 K (for the gas responsible for the profile maxima), which is at
least twice the previous estimates from mm-wave data. Due to the
important role of thermal diffusion in this case (Sect.\ 3.1), the
high temperature in the disk must have strong effects on its shape and
kinematics and can, in particular, explain the presence of gas
evaporation in the outer parts of the disk.

\begin{acknowledgements}
HIFI has been designed and built by a consortium of institutes and
university departments from across Europe, Canada, and the United
States under the leadership of SRON Netherlands Institute for Space
Research, Groningen, The Netherlands, and with major contributions from
Germany, France, and the US.  Consortium members are: Canada: CSA,
U.Waterloo; France: CESR, LAB, LERMA, IRAM; Germany: KOSMA, MPIfR, MPS;
Ireland, NUI Maynooth; Italy: ASI, IFSI-INAF, Osservatorio Astrofisico
di Arcetri- INAF; Netherlands: SRON, TUD; Poland: CAMK, CBK; Spain:
Observatorio Astron\'omico Nacional (IGN), Centro de Astrobiolog\'{\i}a
(CSIC-INTA). Sweden: Chalmers University of Technology - MC2, RSS \&
GARD; Onsala Space Observatory; Swedish National Space Board, Stockholm
University - Stockholm Observatory; Switzerland: ETH Zurich, FHNW; USA:
Caltech, JPL, NHSC. HCSS / HSpot / HIPE is a joint development (are
joint developments) by the Herschel Science Ground Segment Consortium,
consisting of ESA, the NASA Herschel Science Center, and the HIFI,
PACS, and SPIRE consortia. This work has been partially supported by
the Spanish MICINN, program CONSOLIDER INGENIO 2010, grant ``ASTROMOL"
(CSD2009-00038). R.Sz.\ and M.Sch.\ acknowledge support from grant N203
581040 of the Polish National Science Center. K.J.\ acknowledges the
funding from SNSB. J.C.\ acknowledges funding from MICINN, grant
AYA2009-07304. This research was performed, in part, through a JPL
contract funded by the National Aeronautics and Space Administration.
\end{acknowledgements}

\bibliographystyle{aa}

\appendix

\section{Observational results (in online version only)}

In this Appendix we show the full DSB spectra obtained in our
program. USB and LSB frequencies are indicated in the figures. We also
show the frequencies of some remarkable molecular lines, which does not
mean that these lines have been detected in the corresponding
sources. Depending of the line widths and S/N ratio, the spectral
resolutions were averaged to channel widths ranging between 0.5 and 1.5
\kms\ (except for the narrow lines of the Red Rectangle, for which we
adopted a resolution of 0.2 \kms). Selected lines are shown in more
detail (and in {\em LSR} velocity scale) in Figs.\ 1 to 14.

\begin{figure*}
{\hspace{1cm}\vspace{0cm}\resizebox{17cm}{!}{ 
\includegraphics{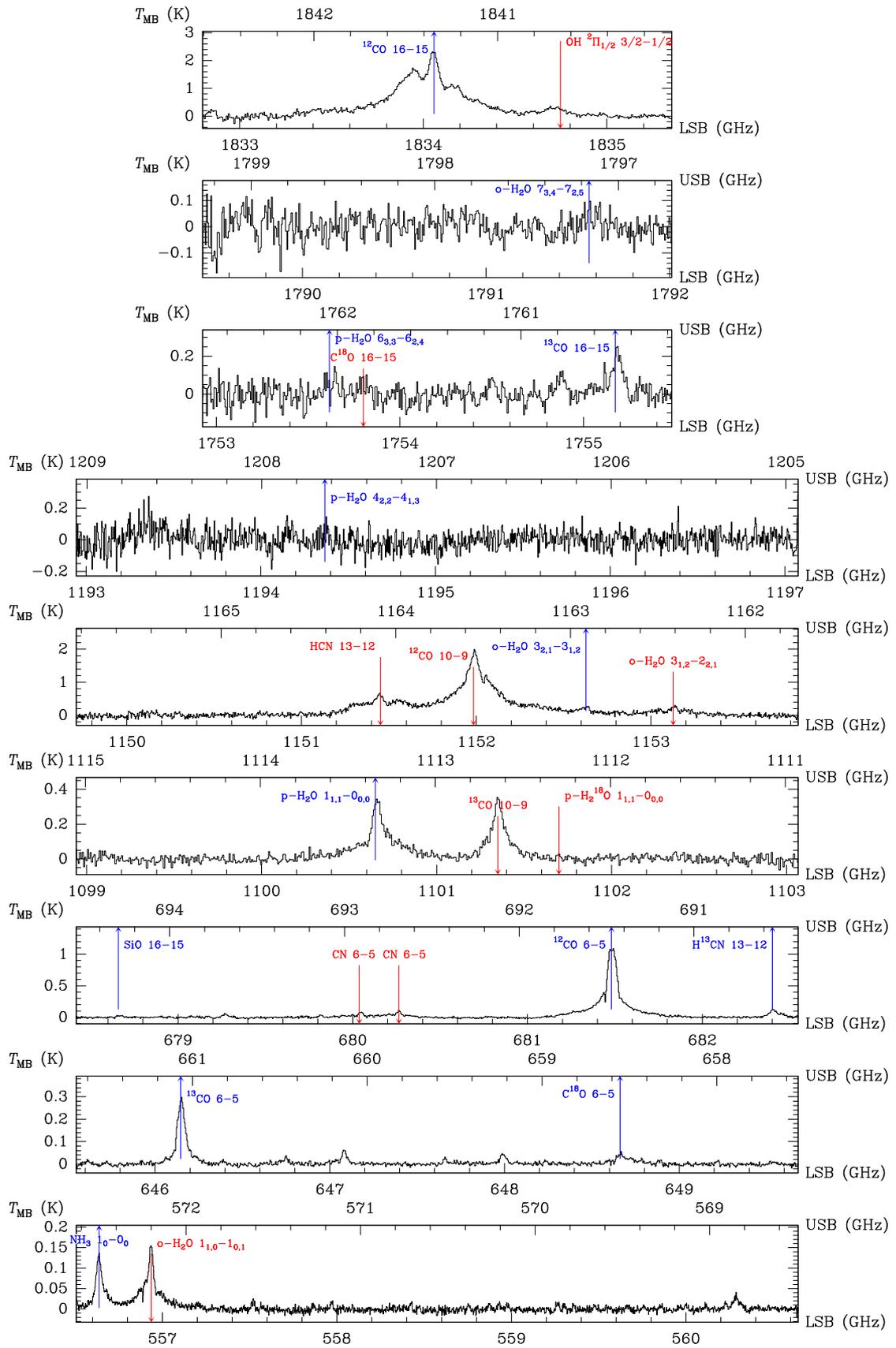}
}}
\caption{Full-band spectra obtained in CRL\,618. We show $T_{\rm mb}$
vs.\ upper- and lower-side band frequencies. Frequencies of some
relevant lines are indicated (which does not mean that they are
detected in this source). All frequencies are corrected for Doppler
shifts due to the relative velocity of the source and therefore
correspond to rest frequencies. Continuum has been subtracted in all
the spectra presented in this paper. }
\end{figure*}

\begin{figure*}
 {\hspace{1cm}\vspace{0cm}\resizebox{17cm}{!}{ 
\includegraphics{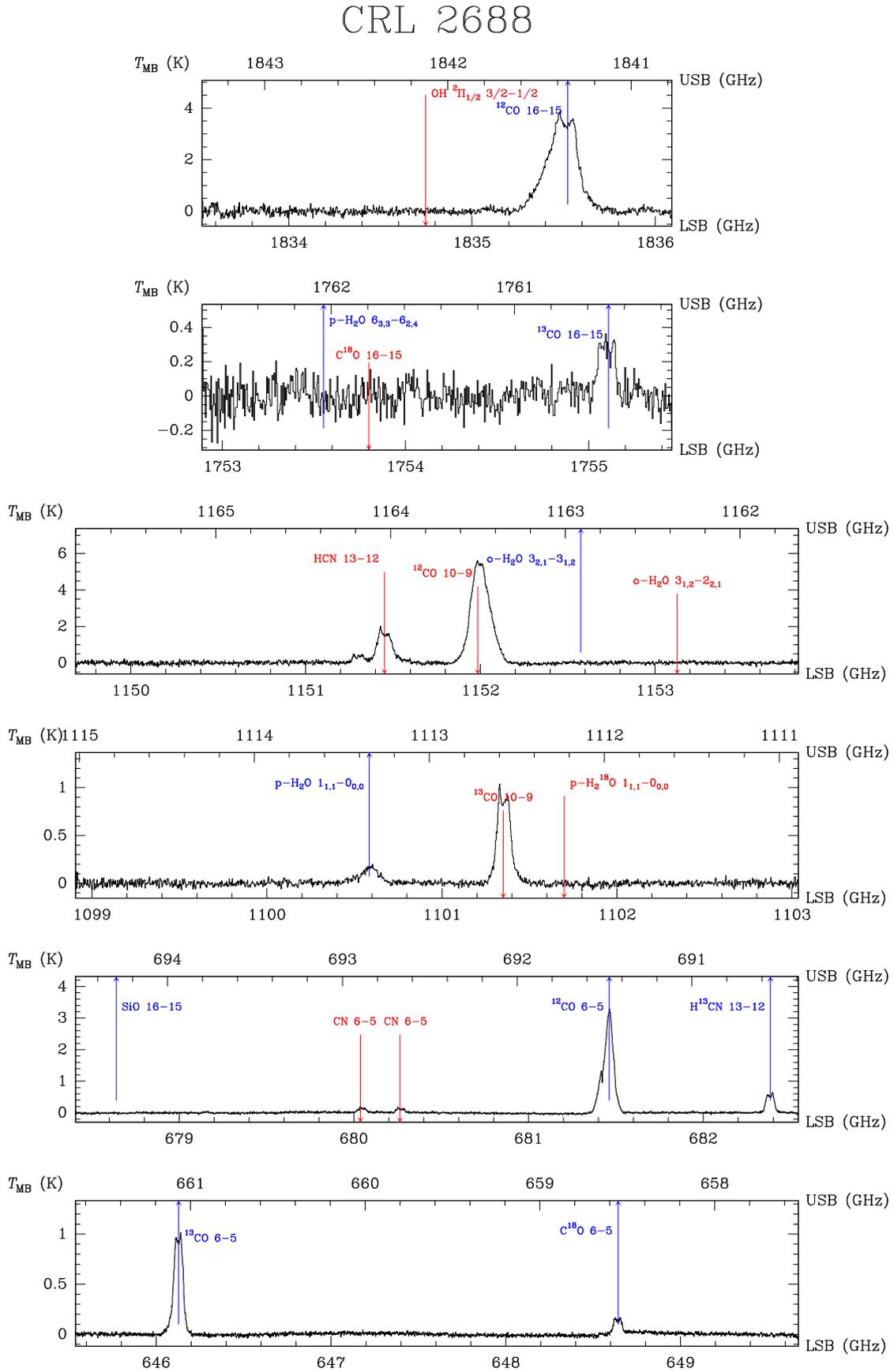}
}}
\caption{Full-band spectra obtained in CRL\,2688; see caption of Fig.\ A.1.
} 
\end{figure*}

\begin{figure*}
{\hspace{1cm}\vspace{0cm}\resizebox{17cm}{!}{ 
\includegraphics{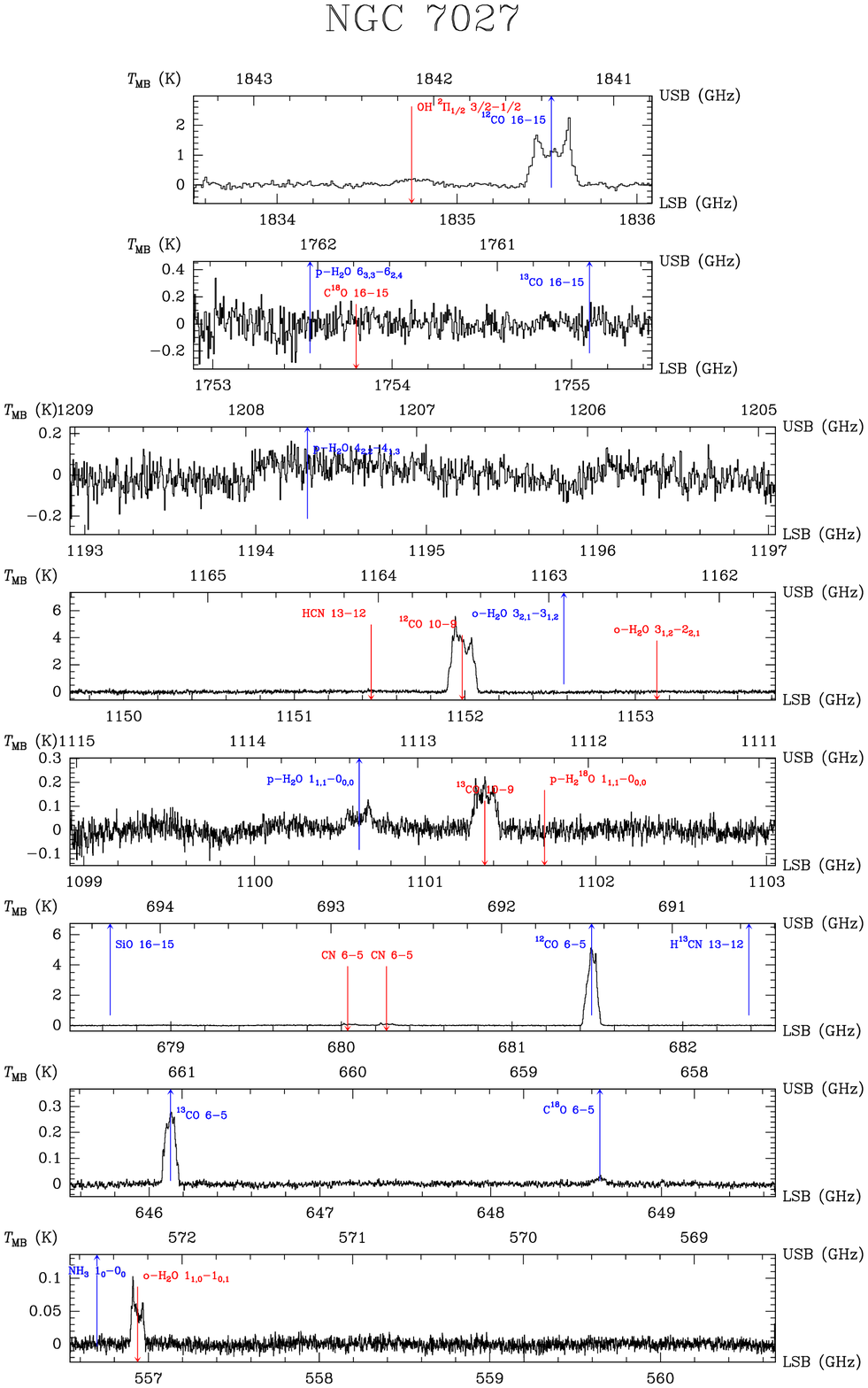}
}}
\caption{Full-band spectra obtained in NGC\,7027; see caption of Fig.\ A.1.
} 
\end{figure*}

\begin{figure*}
{\hspace{1cm}\vspace{0cm}\resizebox{17cm}{!}{ 
\includegraphics{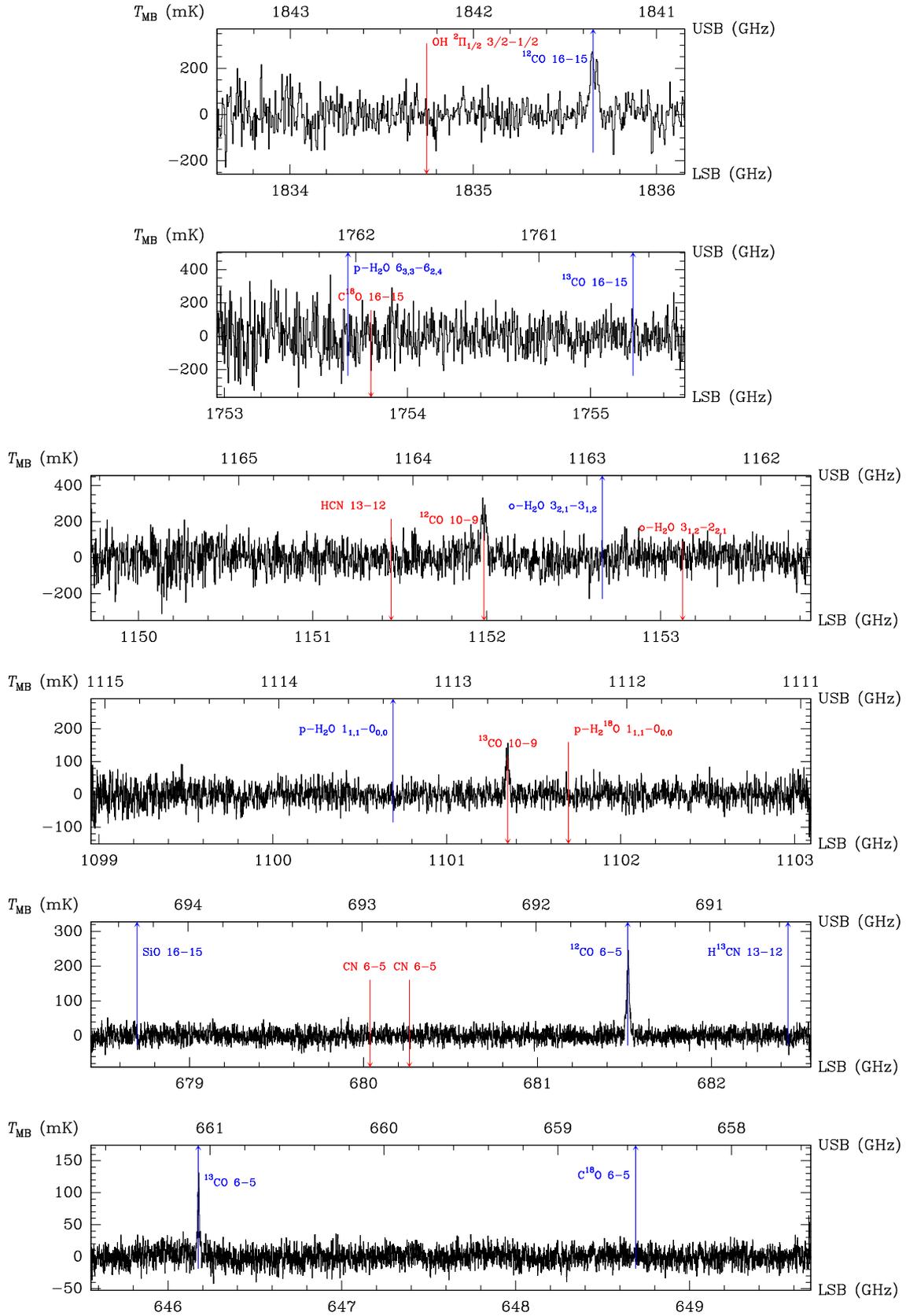}
}}
\caption{Full-band spectra obtained in the Red Rectangle; see caption of
  Fig.\ A.1.
} 
\end{figure*}

\begin{figure*}
{\hspace{1cm}\vspace{0cm}\resizebox{17cm}{!}{ 
\includegraphics{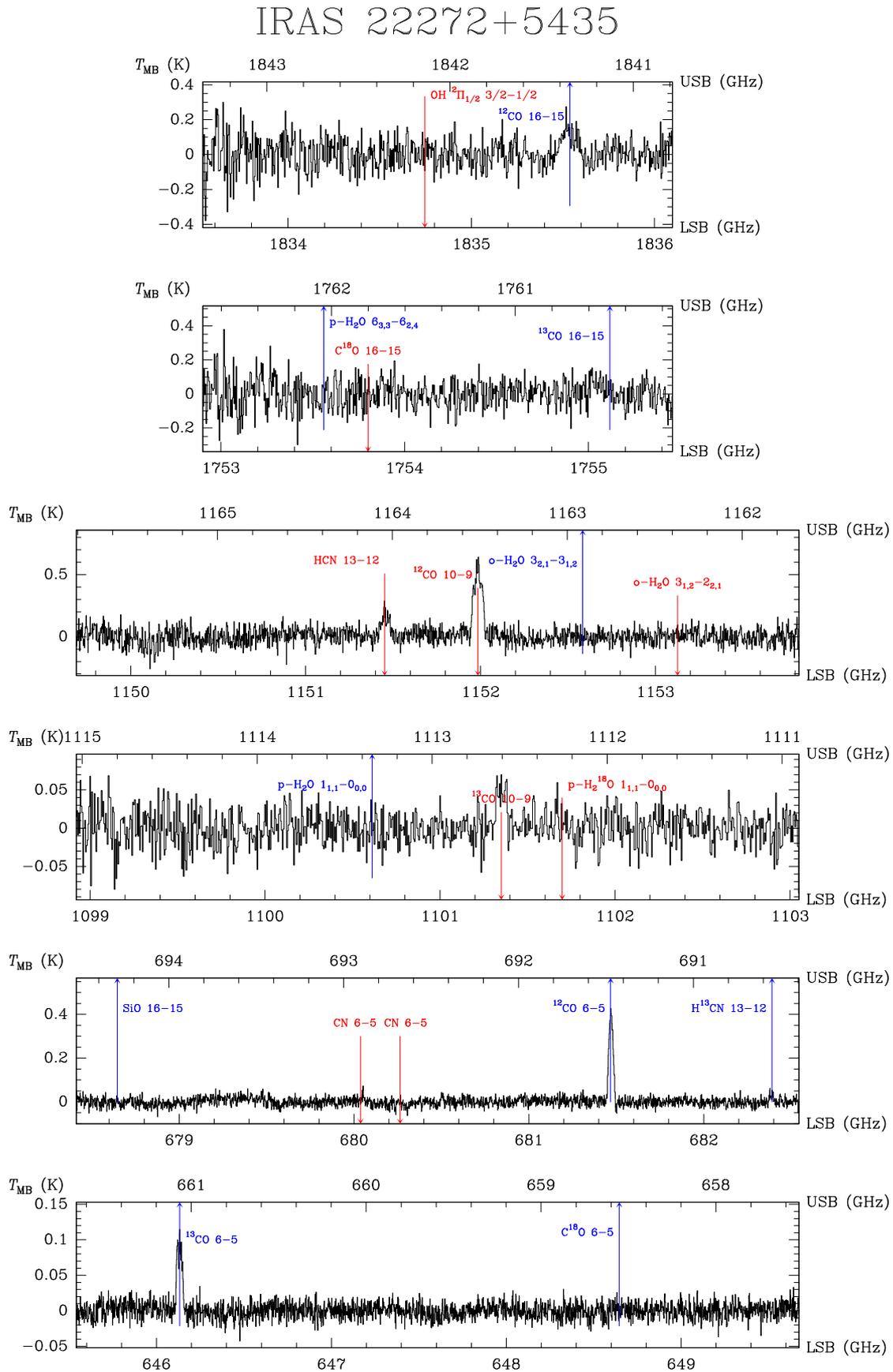}
}}
\caption{Full-band spectra obtained in IRAS\,22272+5435; see caption of
  Fig.\ A.1.
} 
\end{figure*}

\begin{figure*}
{\hspace{1cm}\vspace{0cm}\resizebox{17cm}{!}{ 
\includegraphics{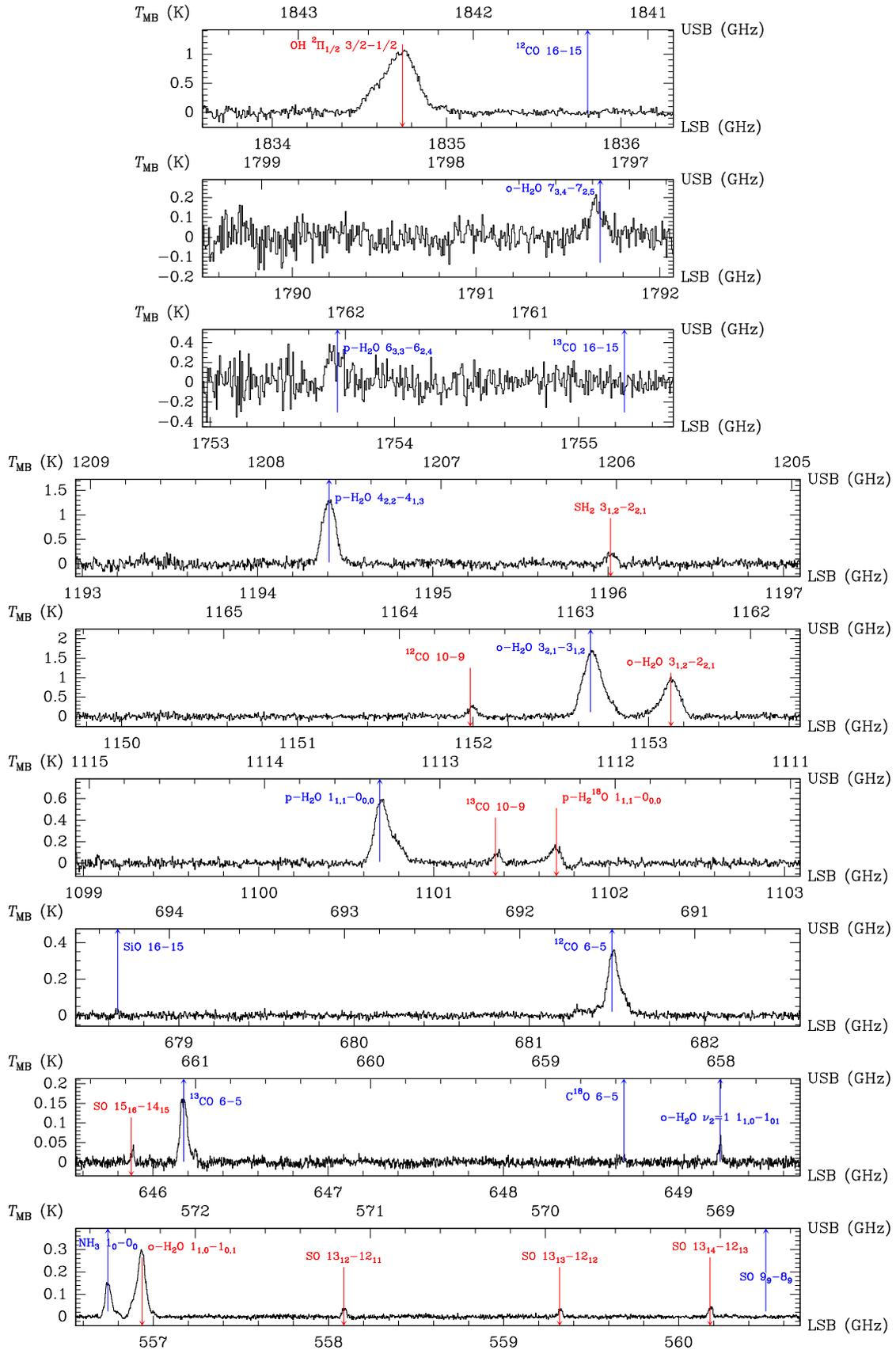}
}}
\caption{Full-band spectra obtained in OH\,231.8+4.2; see caption of
  Fig.\ A.1.
} 
\end{figure*}

\begin{figure*}
{\hspace{1cm}\vspace{0cm}\resizebox{17cm}{!}{ 
\includegraphics{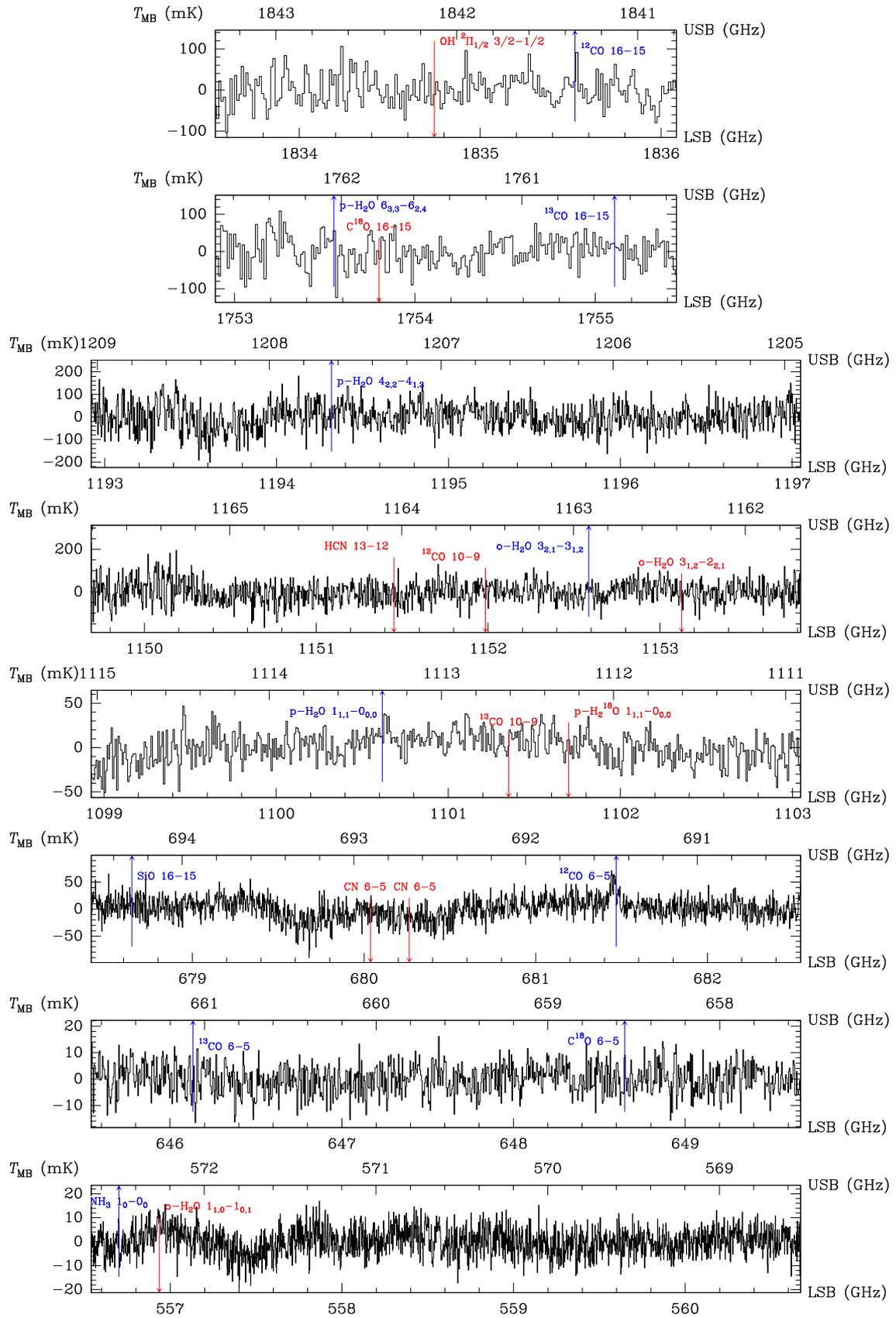}
}}
\caption{Full-band spectra obtained in the Boomerang Nebula; see caption of
  Fig.\ A.1.
} 
\end{figure*}

\begin{figure*}
{\hspace{1cm}\vspace{0cm}\resizebox{17cm}{!}{ 
\includegraphics{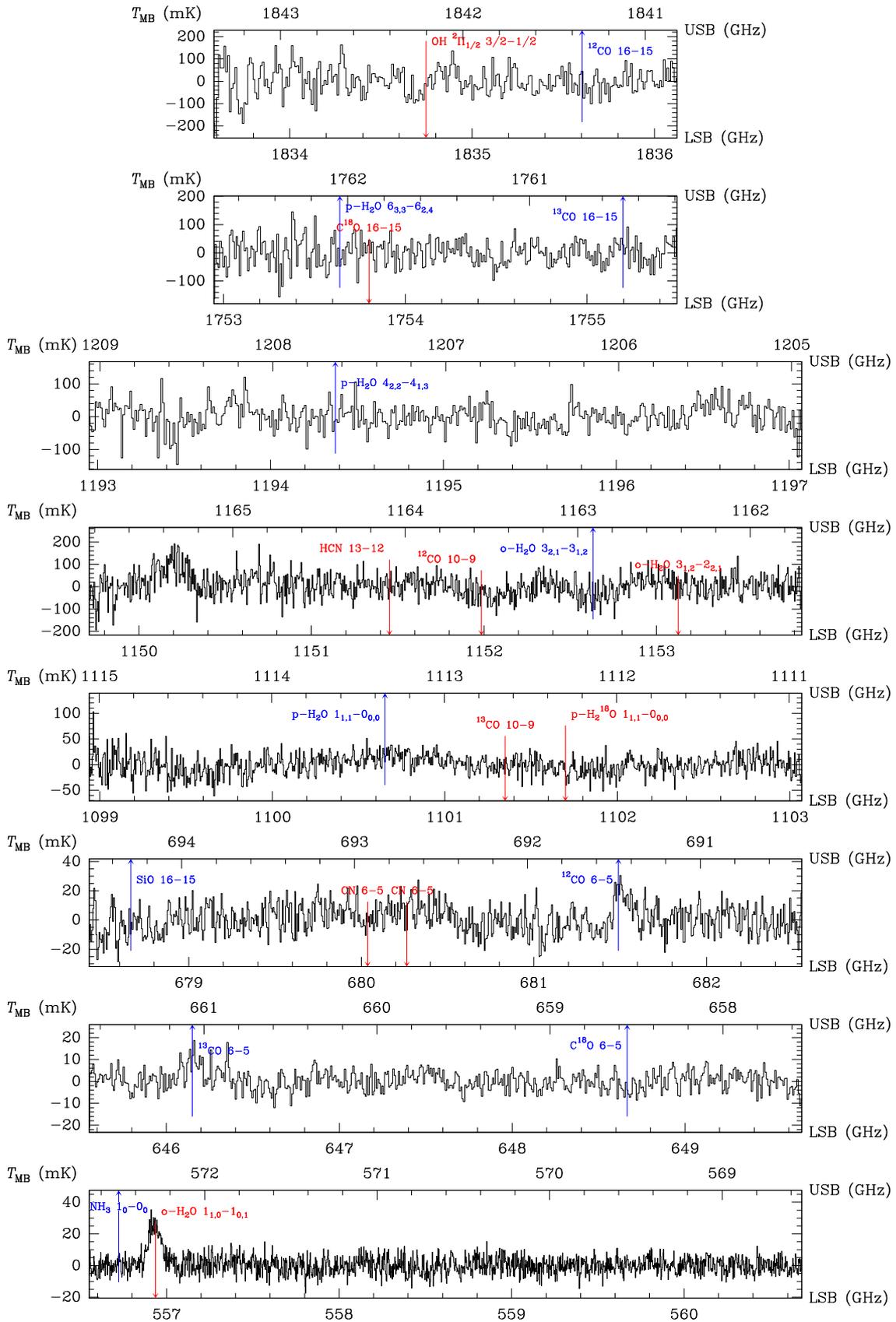}
}}
\caption{Full-band spectra obtained in Frosty Leo; see caption of
  Fig.\ A.1.
} 
\end{figure*}

\begin{figure*}
{\hspace{1cm}\vspace{0cm}\resizebox{17cm}{!}{ 
\includegraphics{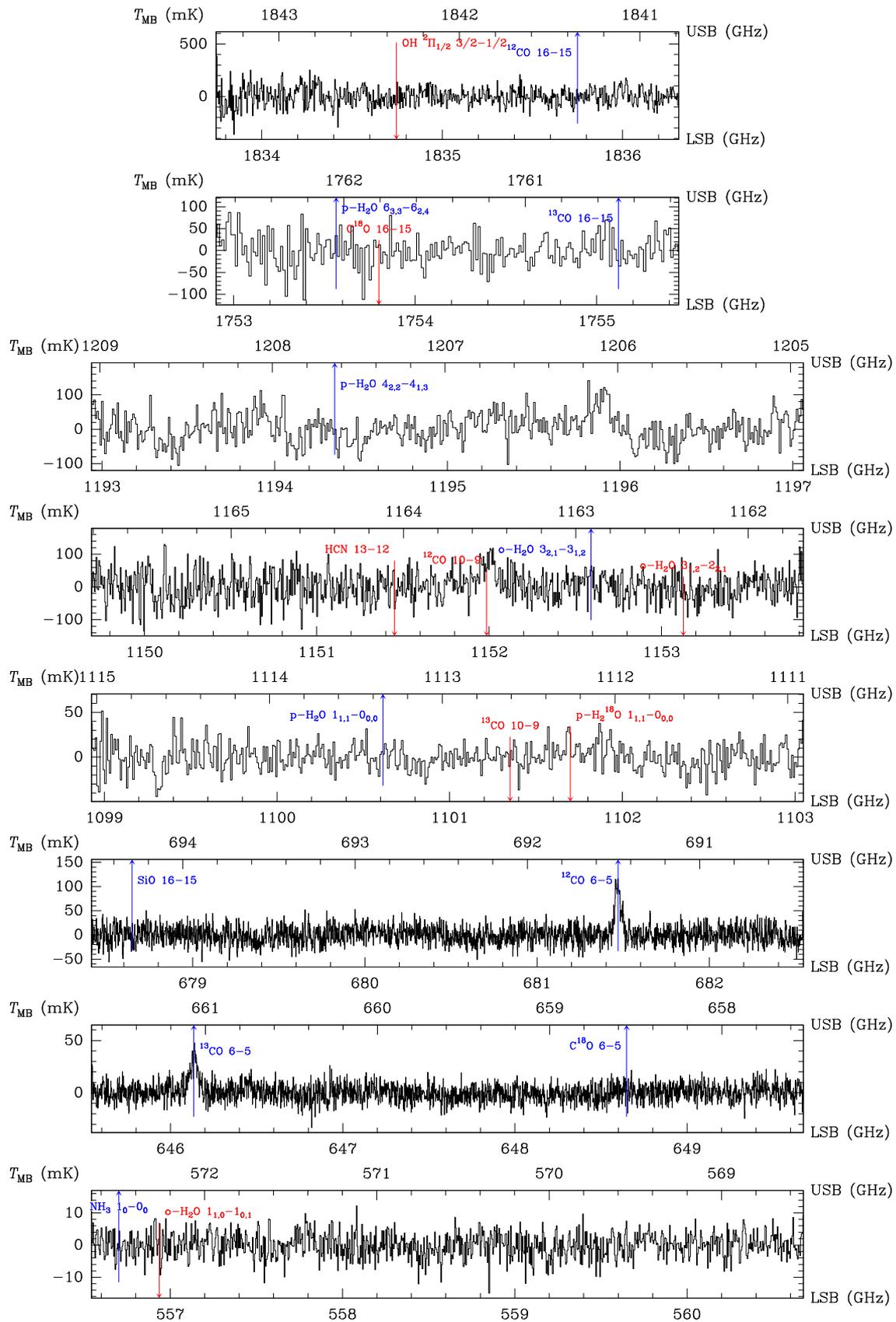}
}}
\caption{Full-band spectra obtained in IRAS\,17436+5003; see caption of
  Fig.\ A.1.
} 
\end{figure*}

\begin{figure*}
{\hspace{1cm}\vspace{0cm}\resizebox{17cm}{!}{ 
\includegraphics{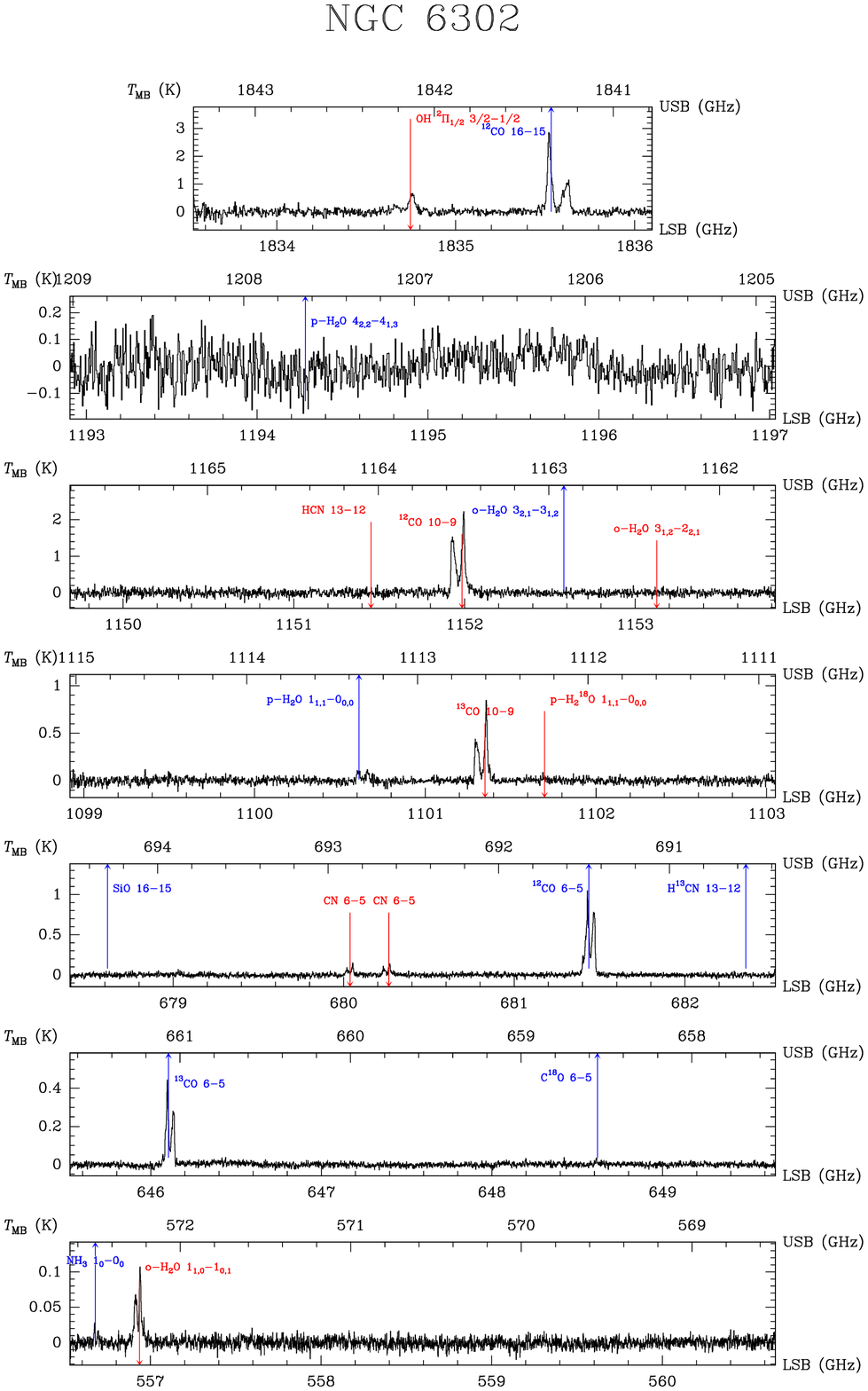}
}}
\caption{Full-band spectra obtained in NGC\,6302; see caption of
  Fig.\ A.1.
} 
\end{figure*}

\end{document}